\documentclass{elsarticle}
\usepackage{lineno,hyperref}

\usepackage[utf8]{inputenc}
\usepackage{bm}
\usepackage{amsmath}
\usepackage{appendix}
\usepackage{xcolor}
\usepackage{amssymb}
\usepackage{soul}

\usepackage{nomencl}
\makenomenclature
\journal{Journal of \LaTeX\ Templates}

\begin{document}

\nomenclature{$A$}
	{strain rate}%
\nomenclature{$\nu$}
	{stoichiometric mass coefficient}
\nomenclature{$(Q)~\bar{Q}$}
	{(dimensionless) heat of combustion}
\nomenclature{$(q)~\bar{q}$}
	{(dimensionless) ratio between the heat transfer and mass transfer rates from the gas to each droplet}
\nomenclature{$(\rho)~\bar{\rho}$}
	{(dimensionless) gas density}
\nomenclature{$(x_i)~\bar{x}_i$}
	{(dimensionless) physical space coordinates}
\nomenclature{$(u)~\bar{u}$}
	{(dimensionless) gas velocity}
\nomenclature{$(\omega)~\dot{\omega}$}
	{(dimensionless) specific chemical reaction rate }
\nomenclature{$(L_v)~\bar{L}_v$}
	{(dimensionless) latent heat}
\nomenclature{$(S_v)~\bar{S}_v$}
	{(dimensionless) source of mass due to the presence of droplets in the flow field}
\nomenclature{$(K)~\bar{K}$}
	{(dimensionless) conductivity}
\nomenclature{$(D)~\bar{D}$}
	{(dimensionless) diffusion coefficient}
\nomenclature{$\alpha_O$}
	{oxidant parameter}
\nomenclature{$(Y_O)~\bar{Y}_O$}
	{(dimensionless) oxidant mass fraction}
\nomenclature{$(Y_{F})~\bar{Y}_F$}
	{(dimensionless) fuel mass fraction}
\nomenclature{$(\Theta)~\bar{T}$}
	{(dimensionless) gas temperature}
\nomenclature{$c_p$}
	{specific heat}
\nomenclature{$\bar{l}_c$}
	{reference distance}
\nomenclature{$\bar{u}_c$}
	{reference axial velocity}
\nomenclature{$\beta$}
	{vaporisation function}
\nomenclature{$\xi$}
	{generic variable}
\nomenclature{$(u_l)~\bar{u}_l$}
	{(dimensionless) droplet velocity}
\nomenclature{$(T_l)~\bar{T}_l$}
	{(dimensionless) droplet temperature}
\nomenclature{$(\rho_l)~\bar{\rho}_l$}
	{(dimensionless) droplet density}	
\nomenclature{$(V_l)~\bar{V}_l$}
	{(dimensionless) liquid volume}	
\nomenclature{$(a)~\bar{a}$}
	{(dimensionless)droplet radius}
\nomenclature{$(g_i)~\bar{g}_i$}
	{(dimensionless) drag force}	
\nomenclature{$(n_l)~\bar{b}_l$}
	{(dimensionless) liquid droplet number density}	

\printnomenclature









\bibliographystyle{elsarticle-num}
\begin{frontmatter}


\title{A generalized spray-flamelet formulation by means of a monotonic variable}


\author[1]{Daniela de Oliveira Maionchi\corref{mycorrespondingauthor}}
\cortext[mycorrespondingauthor]{Corresponding author}
\ead{dmaionchi@fisica.ufmt.br}

\author[2]{Fabio Pereira dos Santos}
\author[3]{Josu\'{e} Melguizo-Gavilanes}
\author[4]{Max Akira Endo Kokubun}
\ead{max@xal.no}

\address[1]{Instituto de F\'isica , Universidade Federal de Mato Grosso - UFMT, 78060-900, Cuiab\'a-MT, Brazil}
\address[2]{Departamento de Engenharia Qu\'imica, Universidade Federal do Rio de Janeiro - UFRJ, 21941-909, Rio de Janeiro-RJ, Brazil}
\address[3]{Institut Pprime, Centre National de la Recherche Scientifique, F86962, Futuroscope - Chasseneuil Cedex, France}
\address[4]{Expert Analytics, Oslo, Norway}

\begin{abstract}
The external structure of the spray-flamelet can be described using the Schvab-Zel'dovich-Li\~nan formulation. 
The gaseous mixture-fraction variable as function of the physical space, $Z(x_i)$, typically employed for the description of gaseous diffusion flames leads to non-monotonicity behaviour for spray flames due to the extra fuel supplied by vaporisation of droplets distributed into the flow. 
As a result, the overall properties of spray flames depend not only on $Z$ and the scalar dissipation rate, $\chi$, but also on the spray source term, $S_v$. 
We propose a new general coordinate variable which takes into account the spatial information about the entire mixture fraction due to the gaseous phase and droplet vaporisation. 
This coordinate variable, $Z_C(x_i)$ is based on the cumulative value of the gaseous mixture fraction $Z(x_i)$, and is shown to be monotonic. 
%
%
For pure gaseous flow, the new cumulative function, $Z_C$, yields the well-established flamelet structure in $Z$-space.
In the present manuscript, the spray-flamelet structure and the new equations for temperature and mass fractions in terms of $Z_C$ are derived and then applied to the canonical counterflow configuration with potential flow. 
Numerical results are obtained for ethanol and methanol sprays, and the effect of Lewis and Stokes numbers on the spray-flamelet structure are analyzed.
The proposed formulation agrees well when mapping the structure back to physical space thereby confirming our integration methodology.    
\end{abstract}

\begin{keyword}
\texttt counterflow diffusion flame\sep spray-flamelet\sep cumulative mixture fraction\sep
monotonic variable\sep Schvab-Zel’dovich-Liñan formulation
\end{keyword}

\end{frontmatter}

\section{Introduction}
Spray combustion is present in a variety of industrial technologies, such as diesel engines, gas turbines and liquid-propellant rockets \cite{CHIU2000}. As a result, modelling it is an important subject that has attracted considerable attention in  the scientific community for many decades. \cite{Faeth1983, JENNY2012, Kah2010, MarcMassot2001, SIRIGNANO2014}. 
Unlike gaseous diffusion flames, that are governed by the competition between scalar mixing and chemistry, spray flames are also influenced by evaporation and mass transport of the liquid-fuel into the reaction zone \cite{Franzelli2015}, making it a more complex problem.

Due to the importance of spray flames, several numerical and theoretical investigations have been performed to understand the main physical and chemical processes that govern  spray combustion  as well as their flame structure in different spatial and temporal scales \cite{Faeth1983,Franzelli2015}. 
Regarding numerical investigations, two different approaches can be highlighted: (i) Eulerian Interface Capturing (EIC), and (ii) Particle Tracking (PT) methods. 
For EIC, Changxiao {\it et al} \citep{Shao2018} developed a computational framework that resolves the interface of the dispersed phase for the atomization, evaporation and combustion processes. 
In order to capture the interface, the authors combined level set and ghost fluid methods.  
Their methodology requires that the computational mesh is of the order of the droplet size, which very quickly results in large meshes that demand the use of extensive computational resources.  
In spite of the fact that their methodology permits to obtain very accurate results, their approach can be intractable if a supercomputer is not used. 
Particularly, for three-dimensional problems such as that presented in~\cite{Shao2018}. 
For PT, Large-Eddy Simulations including a two-phase flow model were performed by Irannejad {\it et al} \citep{Irannejad2015};  
the gaseous phase field was solved using a Eulerian framework and the liquid spray phase using a two-way Lagrangian stochastic method. 
The authors considered methanol spray combustion and obtained good agreement with experimental data.
However, similar to \citep{Shao2018}, the simulations performed in \citep{Irannejad2015} also required extensive computational resources.

Theoretical investigations of the spray-flame structure can give important physical insights into the behaviour of spray flames in simple configurations. Such insights can then be used as a basis to understand more complex combustion systems \cite{Franzelli2015}.
Traditionally, the flame structure of laminar gaseous diffusion flames is studied in terms of the gaseous mixture fraction $Z$ \cite{Peters1984}, a passive scalar that is an appropriate variable to analyse the mixing of the reactants (the dominant physical process in these type of flames).

Besides enabling a more computationally efficient solution in composition space compared to the physical-space solution, the mixture fraction concept is widely used in turbulent combustion models, since it allows the turbulent flame to be described in terms of simple one-dimensional elements called flamelets \cite{bookPoinsot}.
Extending this formulation to spray-flames would in principle enable the analysis tools developed for gaseous flames to be applied. 
However, a direct extrapolation of the classical mixture fraction to spray-flames is not possible because $Z$ becomes non-monotonic due to the presence of vaporisation sources \cite{Franzelli2015,Luo2013,Olguin2014,Sanchez2015}; the constraint of monotonicity is required to guarantee that the solution is single-valued. 
In addition to $Z$, other composition spaces have been proposed and analysed in previous studies such as the total mixture fraction \cite{Smith2000,Urzay2013,Vie2013}, and the conserved mixture fraction \cite{Franzelli2015}.
The aforementioned alternatives do not have their monotonicity guaranteed mainly due to differential diffusion and the relative velocity that exists between the liquid and gaseous phases. 
An effective composition variable combining the gaseous mixture fraction and the liquid-to-gas mass ratio was applied to the analysis of counterflow spray-flames in \cite{Franzelli2015}. 
This variable was then employed to derive the governing equations for a spray-flamelet formulation.
Although this formulation was found to reproduce the response of the flame structure to variations in the droplet diameter and strain rate, it required the use of a closure model for the scalar dissipation rate, $\chi$~\cite{Olguin2019}. 

The main objectives of this work are: (i) to propose an alternative monotonic variable that enables the description of spray-flames and apply it to a simple canonical problem (e.g. counter flow configuration) to highlight the methodology. 
This new variable, termed cumulative mixture fraction, $Z_C$, consists of integrating the usual gaseous mixture fraction $Z$ over physical space, $x$, but weighted by a normal distribution. 
It results in an initially increasing function that reaches a plateau, therefore remaining single-valued and guaranteeing monotonicity; 
(ii) to formulate the spray-flamelet equations in $Z_C$-space. An interesting outcome of this formulations is that no extra model is necessary for the scalar dissipation rate, $\chi$, as its dependence is directly obtained from the $Z_C$ equations; and
(iii) to present simulation results, in both $x-$ and $Z_C-$space, for ethanol and methanol droplets.
Special attention is given to the effect of the Lewis and Stokes numbers on the spray-flamelet structure.

This manuscript is organised as follows: in Section 2, we present the Schvab-Zeldovich-Li\~nan formulation for the spray-flamelet model for both the gaseous phase and the liquid droplets. 
%
%
In Section 3, we derive the model's equations in terms of the strictly monotonic cumulative mixture fraction, $Z_C$. 
In Section 4, we present results for the counterflow configuration with potential flow.
Finally, concluding remarks are presented in Section 5.

\section{Physical model}

The governing equations are formulated in an Eulerian framework, assuming steady-state and the low-Mach number limit for the gas phase \cite{Maionchi2013,Maionchi2017}.
For simplicity, infinitely fast chemistry is considered (Burke-Schumann limit), enabling the diffusion flame to be described in terms of the extended Shvab-Zel'dovich model \cite{Burke1928}. 
In analogy to the theory for gaseous flames, the model developed here can be extended by relaxing these considerations.

A single global reaction step is used to represent the combustion processes according to
%
$$F + \nu O \to (1 + \nu) P + \Bar{Q},$$
where $\bar{Q}$ is the heat release and the stoichiometric mass coefficient is defined as $\nu = m_O/m_F$.

\subsection{Governing equations in physical space - $x$}
\subsubsection{Gaseous Phase}

The gaseous phase is described in $\bm{x} = \{x_1,x_2,x_3\}$ -space by the following dimensionless conservation equations of mass, momentum, fuel/oxidant mass fractions, and energy \citep{Maionchi2013},
\begin{subequations}
\begin{eqnarray}
&&\frac{\partial}{\partial x_i}(\rho u_i)
= \alpha_0 S_v,
\label{eq01a}
\\
&&\frac{\partial}{\partial x_i}(\rho u_i u_j)
=
\frac{\partial}{\partial x_i}\left(\frac{Pr}{Pe}\mu \frac{\partial u_j}{\partial x_i} \right) -  \frac{\partial p}{\partial x_j}  +  \alpha_0 S_v u_{lj}-  g_j,
\label{eq01b}
\\
&&\frac{\partial}{\partial x_i}(\rho u_i Le_O Y_O)
=
\frac{\partial}{\partial x_i}\left(\frac{\Gamma^\gamma}{Pe} \frac{\partial Y_O}{\partial x_i}\right) + \dot{\omega},
\label{eq01c}
\\
&&\frac{\partial}{\partial x_i}(\rho u_i Le_F Y_F)
=
\frac{\partial}{\partial x_i}\left(\frac{\Gamma^\gamma}{Pe} \frac{\partial Y_F}{\partial x_i}\right) + \dot{\omega} + S_v,
\label{eq01d}
\\
&&\frac{\partial}{\partial x_i}(\rho u_i\Theta)
=
\frac{\partial}{\partial x_i}\left(\frac{\Gamma^\gamma}{Pe}  \frac{\partial \Theta}{\partial x_i}\right) - Q \dot{\omega}+ S_v (\alpha_0 T_l - q),
\label{eq01e}
\end{eqnarray}\label{eq01}
\end{subequations}
where $\bm{u}_l$ is the liquid velocity and $\bm{g}$ the drag force; the remaining symbols are listed in the nomenclature.
The heat transfer from gas to liquid is accounted by the last two terms in the right-hand side of Eq.~(\ref{eq01e}).
The non-dimensional mass fractions and temperature are defined as
\begin{equation}
      Y_O \equiv \frac{\bar{Y}_O}{\bar{Y}_{O \infty}}  , \ \ 
      Y_{F} \equiv  \frac{Le_O \nu \bar{Y_F}}{Le_F Y_{O\infty}}, \ \ 
      \Theta \equiv \frac{Le_O \bar{T}}{Y_{O\infty} \bar{T}_\infty}. \ \ 
\label{eq02}
\end{equation}

The subscripts $-\infty$ and $\infty$ represent the fuel and oxidant ambient conditions, respectively.
Thus, $\bar{Y}_{O \infty}$ and $\bar{Y}_{F-\infty}$ are the oxidant and fuel mass fractions in the incoming streams.
The dimensionless variables and characteristic quantities are given by:
\begin{subequations}
\begin{eqnarray}
           &&\rho = \frac{\bar{\rho}}{\bar{\rho}_{\infty}}, \ 
		   x_i = \frac{\bar{x}_i}{\bar{l}_c}, \ 
		   u_i = \frac{\bar{u}_i}{\bar{u}_c}, \	
		   p = \frac{\bar{p}}{\bar{\rho}_{\infty} \bar{u}_c^2}, \ 
           g_j = \frac{\bar{l}_c}{\bar{\rho}_{\infty} \bar{u}_c^2}\bar{g}_j, \ 
           \alpha_O = \frac{\bar{Y}_{O \infty}}{Le_O \nu}, \ \    
\label{eq03a}\\           
           &&\dot{\omega} = \frac{l_c}{\rho_{\infty} u_c \alpha_O} \bar{\dot{\omega}}, \ \
           Q = \frac{\bar{Q}}{\nu c_p \bar{T}_{\infty}},  \ \
           q = \frac{\bar{q}}{\nu c_p \bar{T}_{\infty}},  \ \
           S_v = \frac{l_c}{\rho_{\infty} \bar{u}_c \alpha_O}\bar{S}_v.
\label{eq03b}
\end{eqnarray}\label{eq03}
\end{subequations}


The Peclet and Prandtl numbers are  $Pe = \bar{l}_c \bar{u}_c \bar{\rho}_{\infty} c_p/\bar{K}_{\infty}$ and $Pr = c_p \bar{\mu}_{\infty}/\bar{K}_{\infty}$, respectively, 
whereas the Lewis number is $Le_j = \bar{K}_\infty /(\bar{\rho}_\infty c_p \bar{D}_j)$ for $j = F,O$ (fuel and oxidant, respectively).
The source of mass, $S_v$, is zero on the oxidant side because the droplets are injected only along the incoming fuel stream.
The heat and mass diffusion transport properties are considered to be temperature-dependent, such that $\bar{K}/\bar{K}_{\infty} = \bar{\rho} \bar{D}/\bar{\rho}_{\infty} \bar{D}_{\infty}= \Gamma^\gamma$, with $\Gamma = \Theta/\Theta_{\infty}$ and $\gamma \ne 0$. 

Defining the mixture fraction and the excess of enthalpy as
\begin{equation}
Z \equiv \frac{1+ Y_{F} - Y_O}{1+\phi} , \qquad
H = \frac{\Theta +  (Q-1) Y_{F} + Y_O}{1+\phi},\label{eq04}
\end{equation}
and equating species and energy, Eqs.~(\ref{eq01c})-(\ref{eq01e}), yields the Schvab-Zel'dovich-Li\~nan formulation \citep{Linan1974,Linan1993,Linan2001,Maionchi2017},
\begin{subequations}
\begin{eqnarray}
&&\frac{\partial}{\partial x_i}\left(\rho u_i \int_0^Z L(Z) dZ \right)
=
\frac{\partial}{\partial x_i}\left(\frac{\Gamma^\gamma}{Pe}\frac{\partial Z}{\partial x_i}  \right) + \frac{S_v}{1+\phi},\label{eq05a}\\
&&\frac{\partial}{\partial x_i} \left(\rho u_i H_N \right)
=
\frac{\partial}{\partial x_i}\left(\frac{\Gamma^\gamma}{Pe}\frac{\partial H}{\partial x_i} \right) +\frac{S_h}{1+\phi}, \ \ H_N= H + \int_0^Z N(Z) dZ,
\label{eq05b}
\end{eqnarray}\label{eq05}
\end{subequations}
where $S_h = \left(Q +\alpha_0T_l - q - 1 \right)S_v$ is the modified heat source, and $\phi = \nu Le_O \bar{Y}_{F_{-\infty}}/ Le_F\bar{Y}_{O_{\infty}}$ is the mixture strength.

%
%
%
The functions $L(Z)$ and $N(Z)$ are given by
\begin{subequations}
\begin{eqnarray}
  &&L =
      \left\{
        \begin{array}{ll}
           Le_{O}, & \text{for }\quad Z < Z_f, \\  
           Le_F,   & \text{for }\quad Z > Z_f,
        \end{array}
      \right.
\label{eq07a}\\
  &&N =
      \left\{
        \begin{array}{ll}
           1-Le_{O}, & \text{for }\quad Z < Z_f, \\  
           (1 -Le_F)(1 - Q),   & \text{for }\quad Z > Z_f,
        \end{array}
      \right.
\label{eq07b}
\end{eqnarray}\label{eq07}
\end{subequations}
in which quantities at the flame location are denoted by the subscript \textit{f}.
Since the flame is assumed to be infinitely thin, we have that at the stoichiometric plane, $Y_O = Y_{F} =0$, leading to $Z_f = (1+\phi)^{-1}$ and $H_f = \Theta_f (1+ \phi)^{-1}$.

\subsubsection{Liquid Phase}

The set of droplets present in the spray are assumed to be monodisperse, mono-temperature and monokinetic. 
While the spray is considered dilute, and interactions between droplets and secondary break-up have been neglected, droplet-gas relative motion due to the droplets inertia is accounted for. 
%

The conservation equations for the liquid phase include the total mass of the droplets, momentum and energy, which are respectively given by
\begin{subequations}
\begin{eqnarray}
    &&\frac{\partial}{\partial x_i}(f_l \rho_l u_{li})
    =
    -\alpha_O S_v,
    \label{eq08a}
    \\
    &&\frac{\partial}{\partial x_i}(f_l \rho_l u_{li} u_{lj})
    =
    -\alpha_O S_v u_{lj} + g_i,
    \label{eq08b}
    \\
    &&\frac{\partial}{\partial x_i}(f_l \rho_l u_{li} T_l)
    =
    -\alpha_O S_v\left[T_l -  \tilde{c} \nu (q+ L_v)\right],
    \label{eq08c}
\end{eqnarray}\label{eq08}
\end{subequations}
where $\Tilde{c} = c_p/c_l$ and $f_l = n_l V_l$ is the liquid volume fraction.

The dimensionless variables and characteristic quantities are 
\begin{subequations}
\begin{eqnarray}
           &&V_l = \frac{\bar{V}_l}{\bar{a}_0^3} =\frac{4\pi}{3} \left(\frac{\bar{a}}{\bar{a}_0}\right)^3 = \frac{4\pi}{3}a^3, \ \
		   \rho_l = \frac{\bar{\rho}_l}{\bar{\rho}_{\infty}}, \ \ 
		   n_l= \bar{a}_0^3 \bar{n}_l, \ \ 
\label{eq9a}           \\
		   &&T_l = \frac{\bar{T}_l}{\bar{T}_{\infty}}, \ \
		   t_c = \frac{\bar{a}_0^2}{\bar{\alpha}_{\infty}}=\frac{\bar{l}_c}{\bar{u}_c}, \ \
		   L_v = \frac{\bar{L}_v}{\nu c_p \bar{T}_{\infty}}, \ \
		   \bm{u}_l = \frac{\bar{\bm{u}}_l}{(\bar{l}_c/\bar{t}_c)}.
\label{eq9b}           
\end{eqnarray} \label{eq09}
\end{subequations}

Equations~(\ref{eq08a}) and (\ref{eq08b}) can be combined to obtain
\begin{equation}
    f_l \rho_l u_{li}\frac{\partial}{\partial x_i} u_{lj} = g_j.
    \label{eq10}
\end{equation}
The drag force, which accounts for the momentum exchange between gas and liquid phases, is given by $g_j = f_l \rho_l(u_j-u_{lj})/(a^2 St)$ \cite{Jackson1997}, where $St = t_s^0 / t_c$ is the Stokes number; $t_s^0 = \bar{\rho}_l \bar{a}_0^2 / (72 \bar{\mu}_{\infty})$ is the Stokes time for droplets with initial radius $\bar{a}_0$, or alternatively, the characteristic time for the liquid phase to adjust to changes in the surrounding flow field \cite{Klein2003}.
Equation (\ref{eq10}) can then be rewritten to yield
\begin{equation}
    u_{li}\frac{\partial u_{lj}}{\partial x_i} = \frac{u_j-u_{lj}}{a^2 St}.
    \label{eq11}
\end{equation}

In the limit of small Stokes number, i.e., $St\ll1$, it can be shown that the liquid phase velocity can be asymptotically described as a function of the gas velocity~\cite{Maxey1987}.
In this case, an algebraic relation for $\bm{u}_l$ can be derived which avoids the computation of the momentum equation,  Eq. (\ref{eq08b}), for the liquid phase
\begin{equation}
    u_{lj} = u_j - a^2 St \left(u_i \frac{\partial u_j}{\partial x_i}\right),
    \label{eq12}
\end{equation}
where terms of $\mathcal{O}(St)$ and higher are neglected.
Note that the effective Stokes number, $a^2 St$, depends on the square of the droplet radius, $a^2$.
However, since from the governing equations normalization $a < 1$, the effective Stokes number $a^2 St$ is always guaranteed to be small.

The model for the droplets motion assumes Stokes flow. 
Put differently, the particle-based Reynolds number, $Re_l$, must be smaller than $1$, i.e., $Re_l = 2\bar{\rho}_l \bar{u}_c \bar{a}_0/\bar{\mu}_{\infty} = Re(2\bar{a}_0/\bar{l}_c) < 1$. 
It can be shown that the aforementioned condition is met for Stokes numbers
\begin{equation}
S_t < \frac{\bar{\rho}_l / \bar{\rho} }{18 Re}.
\label{eq13}
\end{equation}

Using representative gas-phase Reynolds numbers, $Re \sim 500$, and mass densities ratios, $\bar{\rho}_l / \bar{\rho} \sim 1000$, from counterflow experiments \cite{niemann2015} our droplets motion model is expected to be valid for $St \lessapprox 0.1$.

\subsection{General spray-flamelet structure}

The spray-flamelet equations follow the formulation derived in \cite{Peters1984} for counterflow gaseous flames, but accounting for a vaporisation source term due to the presence of droplets.
We perform a coordinate transformation $(x_1,x_2,x_3)\longrightarrow(\xi(x_1,x_2,x_3),\xi_2,\xi_3)$.
This new coordinate system is attached to the flame element, with $\xi$ being the coordinate normal to the flame, and $\xi_2,\xi_3$ mutually orthonormal tangential components.
In this coordinate system, the derivatives along the $\xi$-direction are much larger than in the  $\xi_2-$ and $\xi_3-$directions, which yields for mass, momentum, mixture fraction and energy conservation: 
\begin{subequations}
\begin{eqnarray}
    &&\sqrt{\frac{\chi}{2D}}\frac{d}{d\xi}(\rho u_i) 
    = \alpha_0 S_v,    \label{eq14a}\\
    &&\sqrt{\frac{\chi}{2D}} \frac{d}{d\xi} (\rho u_i u_j)
= \mu \frac{Pr}{Pe} \sqrt{\frac{\chi}{2D}}  \frac{d^2u_j}{d\xi^2} +\Sigma_\xi^*\frac{du_j}{d\xi} - J_j\frac{dp}{d\xi}+\alpha_0 S_v u_{lj} -  g_j, \nonumber\\   \label{eq14b}\\
    &&\sqrt{\frac{\chi}{2D}} \frac{d}{d\xi}\left(\rho u_i \int L(Z)dZ \right) 
    = \frac{\rho \chi}{2}\frac{d^2Z}{d\xi^2}+ \Sigma_\xi^\dagger \frac{dZ}{d\xi}+ \frac{S_v}{1+\phi},   \label{eq14c}\\
     &&\sqrt{\frac{\chi}{2D}} \frac{d}{d\xi}\left(\rho u_i H_N\right)
    = \frac{\rho \chi}{2}\frac{d^2H}{d\xi^2}+ \Sigma_\xi^\dagger \frac{dH}{d\xi}+ \frac{S_h}{1+\phi},    \label{eq14d}
\end{eqnarray}\label{eq14}
\end{subequations}
where
\begin{equation}
    \Sigma_\xi^* =  \frac{1}{2}\frac{Pr}{Pe}\mu  \frac{d}{d\xi}\left(\frac{\chi}{2D}\right) + \frac{\chi}{2D}\frac{d}{d\xi} \left(\frac{Pr}{Pe}\mu\right),
    \ \ \ 
    \Sigma_\xi^\dagger = \frac{1}{2}\frac{\Gamma^\gamma}{Pe}\frac{d}{d\xi}\left(\frac{\chi}{2D}\right)+  \frac{\chi}{2D}\frac{d}{d\xi}\left(\frac{\Gamma^\gamma}{Pe}\right),
    \label{eq15}
\end{equation}
are generalised fluxes, and
\begin{equation}
\chi = 2D\frac{\partial\xi}{\partial x_i}\frac{\partial\xi}{\partial x_i},
\label{eq16}
\end{equation}
is the generalised scalar dissipation rate \cite{Peters1984} and $J_j= \partial\xi / \partial x_j$ \cite{Franzelli2015}.

Finally, the coordinate transformation to $\xi-$space for the liquid phase leads to

\begin{subequations}
\begin{eqnarray}
    &&\sqrt{\frac{\chi}{2D}}\frac{d}{d\xi} (f_l \rho_l u_{li})
    =  -\alpha_O S_v,    \label{eq17a}\\
    &&u_{lj} = u_j - a^2 St \sqrt{\frac{\chi}{2D}}\left(u_i\frac{du_j}{d\xi}\right),
    \label{eq17b}\\
    &&\sqrt{\frac{\chi}{2D}}\frac{d}{d\xi}(f_l \rho_l u_{li} T_l)
    =    -\alpha_O S_v\left[T_l -  \tilde{c} \nu (q + L_v)\right].
    \label{eq17c}
\end{eqnarray}\label{eq17}
\end{subequations}

The set of Eqs (\ref{eq14})--(\ref{eq17}) define the generalised spray-flamelet equations in $\xi$-space, with constant but non-unity Lewis numbers and accounting for temperature and velocity differences for the liquid phase.

%
In the next Section we will show the choice of the function $\xi$ that naturally leads to a monotonic description of the spray-flamelet equations for a counterflow configuration. 

\section{Strictly monotonic Cumulative Mixture Fraction Function $Z_C$}

%

\begin{figure}[h!]
    \centering
    \includegraphics[width=0.75\linewidth]{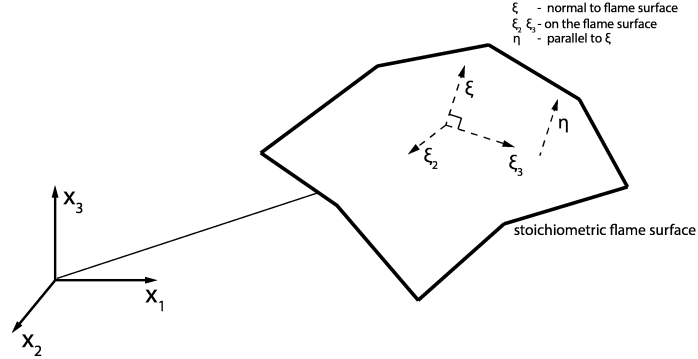}
    \caption{Schematic of the coordinate systems considered: $x_i$ are the laboratory coordinates; $\xi_i$ are the flame-attached coordinates, with $\eta$ parallel to the normal coordinate along the flame, $\xi$.}
    \label{fig0}
\end{figure}


%
The proposed approach is to define a new coordinate $Z_C$ as the integral of the mixture fraction $Z$ along $\eta$, which is an integration variable parallel to $\xi$. Since $Z_C$ is integrated along the parallel direction of $\xi$, we thereby guarantee that it is also parallel to the flame surface. 
This allows us to choose $Z_C$ as the generic variable $\xi$ and to write $d\eta = (\partial \eta/\partial x_i)dx_i$. 
A schematic of the coordinate systems considered in this work is shown in Fig.~\ref{fig0}. 
Note that although $\xi$ and $\eta$ are parallel, they are not the same variable. 
The following coordinate transformation can thus be written:
\begin{equation}
    \xi(\eta) = Z_C(\eta) = \frac{1}{Z_C^T}\int_{-\infty}^{\eta} \frac{e^{-s^2/2}}{\sqrt{2\pi}} Z(s) ds, \ \ \ Z_C^T = \int_{-\infty}^{\infty} \frac{e^{-s^2/2}}{\sqrt{2\pi}} Z(s) ds.
    \label{eq18}
\end{equation}
%

%
In particular, we have that
\begin{equation}
    \frac{dZ_C}{d\eta} = \frac{e^{-\eta^2/2}}{\sqrt{2\pi}} \frac{Z(\eta)}{Z_C^T},
    \label{eq19}
\end{equation}
which enables the calculation of $Z_C^T$ and the conversion back to $\eta$-space
\begin{equation}
    Z_C^T = \frac{\int_{-\infty}^\infty e^{-\eta^2/2}d\eta/\sqrt{2\pi}}{\int_0^1 dZ_C/Z(Z_C)}, \ \
    \eta = Z_C^T \int_0^{Z_C} \sqrt{2\pi} e^{\eta^2/2} \frac{dZ_C'}{Z(Z_C')}.
    \label{eq20}
\end{equation}

Notably, this choice enables us to find a closure relation for the dissipation rate as
\begin{equation}
    \frac{\chi}{2D} = \left(\frac{dZ_C}{dx_i}\right)^2= \frac{\chi_\eta}{2D} \left(\frac{\partial\eta}{\partial x_i}\right)^2, \quad  \frac{\chi_\eta}{2D} = \frac{e^{-\eta^2}}{{2\pi}}\left(\frac{Z(\eta)}{Z_C^T}\right)^2
    \label{eq21}
\end{equation}
Note that this definition differs from \cite{Peters1984} in which $\xi = Z$ is used as independent variable: $\chi_Z = 2D (dZ/dx_i)^2$.
Moreover, it is possible to obtain an expression for $\chi$ defined in Eq.~(\ref{eq21}),
\begin{equation}
    \frac{\chi}{2D} = \left(\frac{dZ_C}{d\eta}\frac{d\eta}{dZ}\frac{dZ}{dx_i}\right)^2 =\left( \frac{1}{Z_C^T} \frac{e^{-\eta^2/2}}{\sqrt{2\pi}}Z(\eta)\right)^2 \left(\frac{d\eta}{dx_i}\right)^2,
    \label{eq21a}
\end{equation}
its dependence on the spatial transformation $d\eta/dx_i$ is readily seen.

Using~(\ref{eq21}) in Eqs.~(\ref{eq14}) yields:
\begin{subequations}
\begin{eqnarray}
    &&\sqrt{\frac{\chi_\eta}{2D}} \frac{\partial\eta}{\partial x_i}\frac{d}{dZ_C}(\rho u_i) 
    = \alpha_0 S_v,    \label{eqp1a}\\
    &&\sqrt{\frac{\chi_\eta}{2D}} \frac{\partial\eta}{\partial x_i}\frac{d}{dZ_C} (\rho u_i u_j)
= \mu \frac{Pr}{Pe} \sqrt{\frac{\chi_\eta}{2D}} \frac{\partial\eta}{\partial x_i}  \frac{d^2u_j}{dZ_C^2} +\Sigma_\xi^*\frac{du_j}{dZ_C} - J_j\frac{dp}{dZ_C}+\alpha_0 S_v u_{lj} -  g_j, \nonumber\\   \label{eqp1b}\\
    &&\sqrt{\frac{\chi_\eta}{2D}} \frac{\partial\eta}{\partial x_i} \frac{d}{dZ_C}\left(\rho u_i \int L(Z)dZ \right) 
    = \rho D \frac{\chi_\eta}{2D} \left(\frac{\partial\eta}{\partial x_i}\right)^2\frac{d^2Z}{dZ_C^2}+ \Sigma_\xi^\dagger \frac{dZ}{dZ_C}+ \frac{S_v}{1+\phi},   \label{eqp1c}\\
     &&\sqrt{\frac{\chi_\eta}{2D}} \frac{\partial\eta}{\partial x_i}\frac{d}{dZ_C}\left(\rho u_i H_N\right)
    = \rho D \frac{\chi_\eta}{2D} \left(\frac{\partial\eta}{\partial x_i}\right)^2 \frac{d^2H}{dZ_C^2}+ \Sigma_\xi^\dagger \frac{dH}{dZ_C}+ \frac{S_h}{1+\phi},    \label{eqp1d}
\end{eqnarray}\label{eqp1}
\end{subequations}
where
\begin{eqnarray}
    &&\Sigma_\xi^* =  \frac{1}{2}\frac{Pr}{Pe}\mu  \frac{d}{dZ_C}\left[\frac{\chi_\eta}{2D} \left(\frac{\partial\eta}{\partial x_i}\right)^2\right] + \frac{\chi_\eta}{2D} \left(\frac{\partial\eta}{\partial x_i}\right)^2\frac{d}{dZ_C} \left(\frac{Pr}{Pe}\mu\right), \nonumber   \\ 
    &&\Sigma_\xi^\dagger = \frac{1}{2}\frac{\Gamma^\gamma}{Pe}\frac{d}{dZ_C}\left[\frac{\chi_\eta}{2D} \left(\frac{\partial\eta}{\partial x_i}\right)^2\right]+  \frac{\chi_\eta}{2D} \left(\frac{\partial\eta}{\partial x_i}\right)^2\frac{d}{dZ_C}\left(\frac{\Gamma^\gamma}{Pe}\right).
    \label{eqpp}
\end{eqnarray}
Similarly, Eqs.~(\ref{eq17}) become
\begin{subequations}
\begin{eqnarray}
    &&\sqrt{\frac{\chi_\eta}{2D}} \frac{\partial\eta}{\partial x_i}\frac{d}{dZ_C} (f_l \rho_l u_{li})
    =  -\alpha_O S_v,    \label{eqp2a}\\
    &&u_{lj} = u_j - a^2 St \sqrt{\frac{\chi_\eta}{2D}} \frac{\partial\eta}{\partial x_i}\left(u_i\frac{du_j}{dZ_C}\right),
    \label{eqp2b}\\
    &&\sqrt{\frac{\chi_\eta}{2D}} \frac{\partial\eta}{\partial x_i}\frac{d}{dZ_C}(f_l \rho_l u_{li} T_l)
    =    -\alpha_O S_v\left[T_l -  \tilde{c} \nu (q + L_v)\right].
    \label{eqp2c}
\end{eqnarray}\label{eqp2}
\end{subequations}

Essentially, this formulation depends on the relation between the locally normal flame coordinate $\eta$ and the physical coordinate $x_i$ through the derivatives $d\eta/d x_i$.
It is worth emphasizing that using the spray-flamelet formulation just derived in multidimensional numerical simulations of turbulent diffusion flames would then entail the implementation of  the $d\eta/d x_i$ relation in each flamelet.

In the next Section, we highlight the main features of this new mathematical framework by considering a planar counterflow configuration.

\section{Results and Discussion}

\begin{figure}[ht]
    \centering
    \includegraphics[scale=0.25]{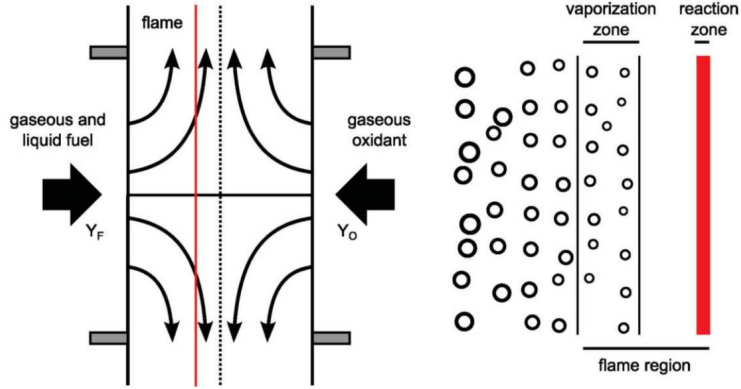}
    \caption{Spray-flamelet model with fuel injected from the left and air from the right side. Left:  Schematic of canonical problem considered. Right: Close up to flame region. Reproduced from \cite{Maionchi2013}.}
    \label{fig1}
\end{figure}

A schematic of the case setup is included in Fig.~\ref{fig1}, whose main assumption is a constant density ($\rho=1$), monodisperse fuel spray.
If we additionally consider $t_c = A^{-1}$ as the characteristic time scale with $A$ being the strain rate, the flow field is described in its dimensionless form as a potential flow $\bm{u} = (-x_1,x_2)$, in physical $x_i$-space i.e., $(x_1, x_2)$.

For this particular configuration, where the flow is aligned with the $x_1$-axis, $d\eta/d x_1 = 1$. 
Furthermore, note that the choice of a potential flow implies that the droplets do not disturb the gaseous flow field (this is analogous to assuming that the liquid volume fraction is negligible, i.e., $n_l V_l = f_l\ll1$) which in turn implies that the drag force $g \sim n_l V_l$ in Eq.~(\ref{eqp1b}) can be neglected. 

Upon applying the simplifications mentioned above, the system of equations for the gaseous phase becomes
\begin{subequations}
\begin{eqnarray}
&&\sqrt{2\pi} e^{x^2/2}\left[x\frac{\Gamma^\gamma}{Pe}\frac{dZ}{dZ_C}-\frac{d}{dZ_C}\left(x \int L(Z)dZ \right)\right] -  \frac{d}{dZ_C}\left[\frac{\Gamma^\gamma}{Pe}\frac{Z}{Z_C^T}\frac{dZ}{dZ_C}\right]
         =\frac{2\pi e^{x^2} S_v}{1+\phi}\frac{Z_C^T}{Z}, \nonumber \\
    \label{eq22a}\\
&& \sqrt{2\pi} e^{x^2/2}\left\{x\frac{\Gamma^\gamma}{Pe}\frac{dH}{dZ_C}-\frac{d}{dZ_C}\left(x H_N\right) \right\}-\frac{d}{dZ_C}\left[\frac{\Gamma^\gamma}{Pe}\frac{Z}{Z_C^T}\frac{dH}{dZ_C}\right] 
            =\frac{2\pi e^{x^2} S_h}{1+\phi}\frac{Z_C^T}{Z}.
    \label{eq22b}
\end{eqnarray}\label{eq22}
\end{subequations} 
where $x=x_1$ only, because the variations along $x_2$ are small.

The boundary conditions for Eqs. (\ref{eq22}) are given by
\begin{equation}
\begin{array}{lll}
   Z = 1, &
   H = [\Theta_{-\infty} + (Q-1) \phi]/(1+ \phi),
   &  \text{ for} \quad Z_C \to 0,  \\
   Z = 0, &
   H = (\Theta_{\infty} + 1)/(1+\phi),
   &  \text{ for} \quad Z_C \to 1.  \label{eq06}
\end{array} 
\end{equation}

Using Eq.~(\ref{eq12}) the velocity of the droplets is
\begin{equation}
    u_l = -x(1 + a^2 St),
    \label{eq23}
\end{equation}
Additionally, the droplets are injected at its boiling temperature $T_B$, which eliminates  the energy equation for the droplets in the governing equations, i.e. Eq.~(\ref{eq17c}).

The model for the vaporisation of droplets is modified from \cite{Maionchi2013} to account for the gas-liquid relative motion.
The source term can then be written as
\begin{equation}
S_v  =  M \lambda_{ef}, \ \ \ M= \frac{3 Le_O \nu }{Y_{O\infty}} \frac{f_l \rho_l}{a^3} (1+a^2 St).    \label{eq24}  
\end{equation}
where $\lambda_{ef}= \lambda(x)/(1+a^2 St)$.
More details are included in Appendix~\ref{ap}.

Two liquid fuels are considered, ethanol (C$_2$H$_6$O) and methanol (CH$_4$O) for the spray in the simulations, whose chemical heat release, latent heat, boiling temperature and mass density (at its boiling temperature) are presented in Table~\ref{tab01}. 
The remaining parameters are: $\gamma=0$, $\bar{Y}_{O \infty} = 0.21$, $\phi = Le_O/Le_F$, $\bar{c}_p = 1.0$ kJ/kg K, $\bar{T}_{-\infty} = 300$ K, $\bar{T}_{\infty} = 400$ K , $\bar{\rho}_{\infty} = \bar{\rho}_{air} = 0.88$ kg/m$^3$ and $f_l = 5 \times 10^{-4}$; Eq.~(\ref{eq24}), yields $M=10$ and $14$ for CH$_4$O and C$_2$H$_6$O, respectively. 
\begin{table}[!h]
\centering
\caption{Fuel properties.}\label{tab01}
\begin{tabular}{lcccccc}
\hline
Fuel    & $\nu$ & $\bar{Q}$ (kJ/g)  & $\bar{l}$ (kJ/g) & $\bar{T_B}$ (K) & $\bar{\rho}_l$ (kg/m$^3$) & M\\
\hline
CH$_4$O      & 1.5   & 22.3 &  1.18  &  338 & 792 & 10\\
C$_2$H$_6$O        & 2.087  & 29.7 &  0.846  &  351 & 789 & 14\\
\hline
\end{tabular}
\end{table}

The following global stoichiometric reactions are used

$$\text{C} \text{H}_4 \text{O} + 1.5 \text{O}_2 \to \text{C} \text{O}_2 + 2 \text{H}_2 \text{O},$$
$$\text{C}_2 \text{H}_6 \text{O} + 3 \text{O}_2 \to 2 \text{C} \text{O}_2 + 3 \text{H}_2 \text{O},$$

The system of integro-differential equations~(\ref{eq22}) were discretized using finite volumes. Adaptive mesh refinement, based on the temperature gradient, was used to ensure adequate resolution of the spray-flame structure.
For the discretization of the diffusive and convective terms a second order central difference scheme and a first order upwind interpolation (to avoid spurious oscillations near the flame) were implemented, respectively. 
%
All simulations were performed using a pseudo-transient approach to better control the numerical stability of the solution \cite{WangC14}. 

To solve Eqs.~(\ref{eq22}) an iterative algorithm was developed using the following methodology. 
First, with an initial condition for $Z$, a first guess for $Z_C^T$ was calculated by Eq.~(\ref{eq18}) using an adaptive trapezoidal integration rule based on the computational mesh in $x$-space (with an initial size of $10^4$ nodes). 
Second, with $Z_C^T$, a first prediction to the solution of the system~(\ref{eq22}) was determined.
Third, a new guess to $Z_C^T$ can be computed with the predicted values of $H$ and $Z$, again, with Eq.~(\ref{eq18}). 
This procedure was repeated until predicted and corrected values for $Z_C^T$, $Z$ and $H$ converged to an $L_1$-norm within $10^{-10}$.

\begin{figure}[h!]
\begin{center}
\resizebox*{\textwidth}{!}{
\includegraphics{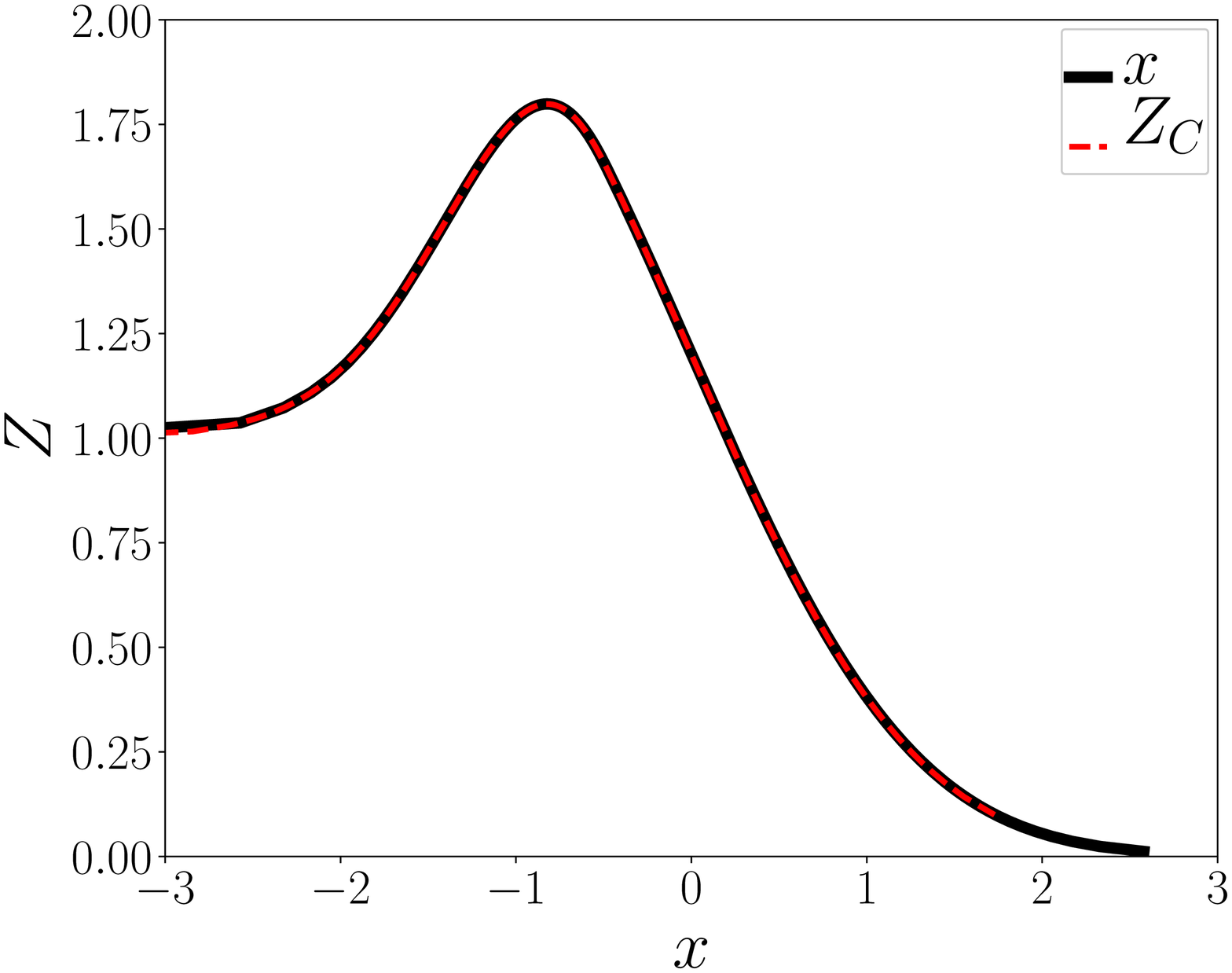}
\includegraphics{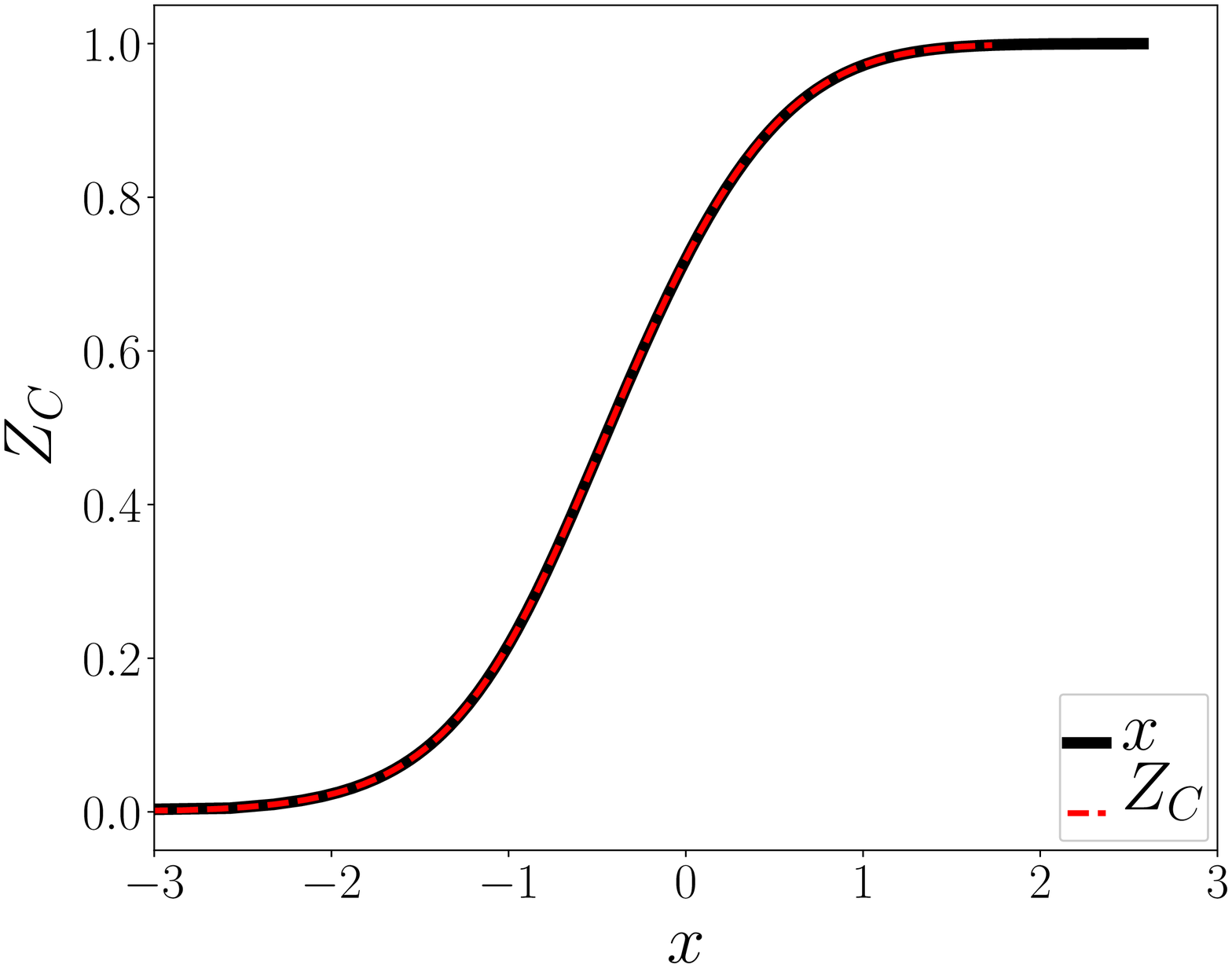}
}
\hspace{0.05\textwidth} (a)\hspace{0.5\textwidth} (b) 
\caption{(a) The mixture fraction $Z$ and (b) the cumulative mixture fraction $Z_C$ of ethanol in terms of physical space $x$.
The generic variable $Z_C$ is a monotonic function of $x$, which is not the case for $Z$ usually adopted in gaseous flows.
The solution in physical space is shown by lines and the corresponding reference solution from the $Z_C$-space formulation converted back to the $x$-space is shown by open circles.}\label{fig02}
\end{center}
\end{figure}

The profiles of $Z$ and $Z_C$ are shown in Fig.~\ref{fig02}. 
The solution in physical space is shown by solid black lines and the corresponding reference solution obtained using the $Z_C$-space formulation mapped to $x$-space is shown by red dashed lines.
The relative deviation between both profiles is less than $10^{-6}$ over the entire domain.
Figure~\ref{fig02}(a) shows that the mixture fraction $Z$, a monotonic variable in purely gaseous flows, is no longer monotonic in a model that includes evaporating droplets in liquid phase.
$Z_C$, on the other hand, remains single-valued   as seen in Fig.~\ref{fig02}(b).


Results for $Le = 1$, and $St = 0$ are presented in subsection~\ref{subsec:unityLe} to highlight the main strengths of the proposed formulation.
Subsections~\ref{subsec:Le_effects} and~\ref{subsec:St_effects} show the effects of  non-unity Lewis numbers, and variation of the Stokes number (i.e. $St < 0.1$) on the spray-flame structure.

\subsection{Unity $Le$ and zero $St$}
\label{subsec:unityLe}

\begin{figure}[hb!]
\begin{center}
\resizebox*{\textwidth}{!}{
\includegraphics{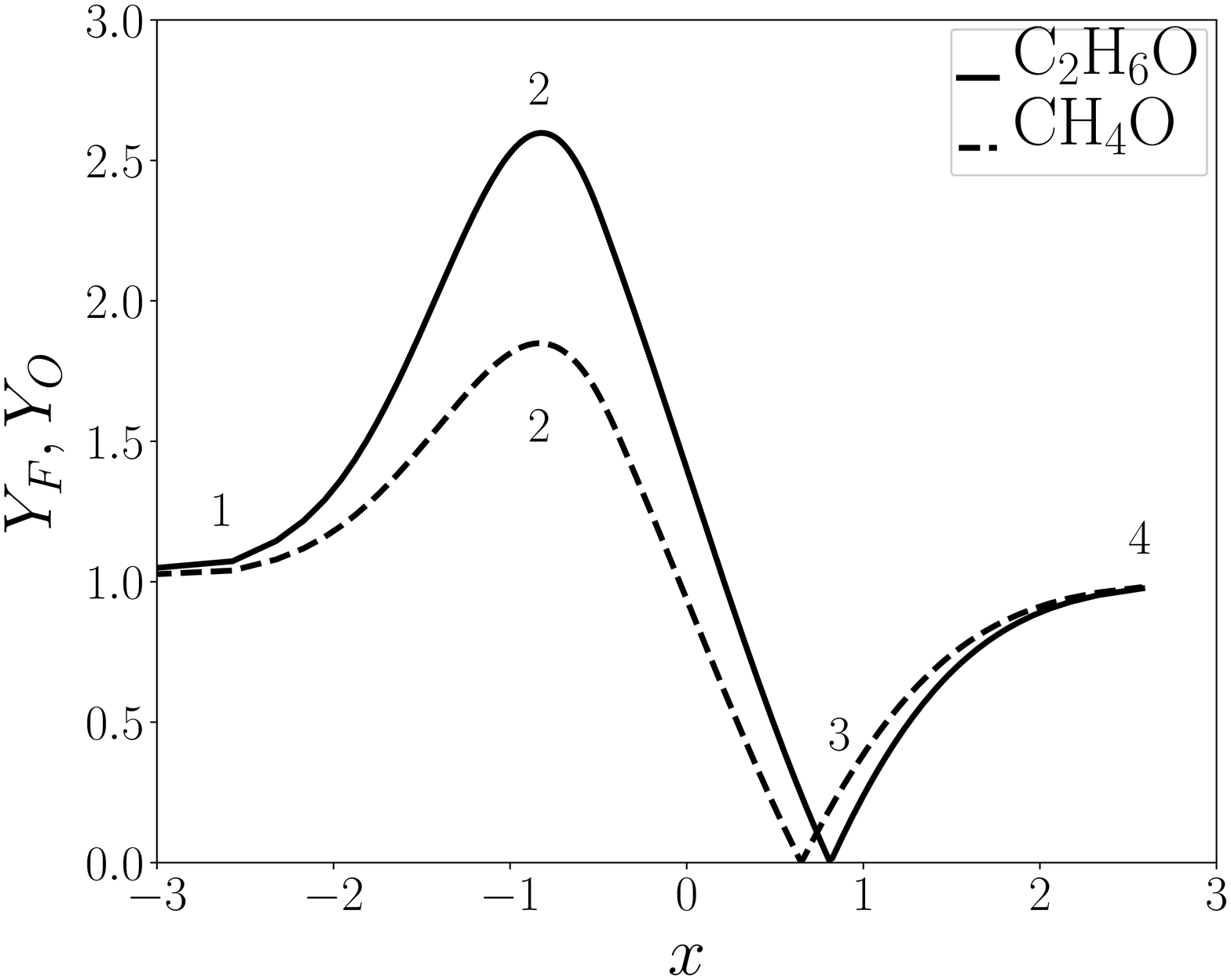}
\includegraphics{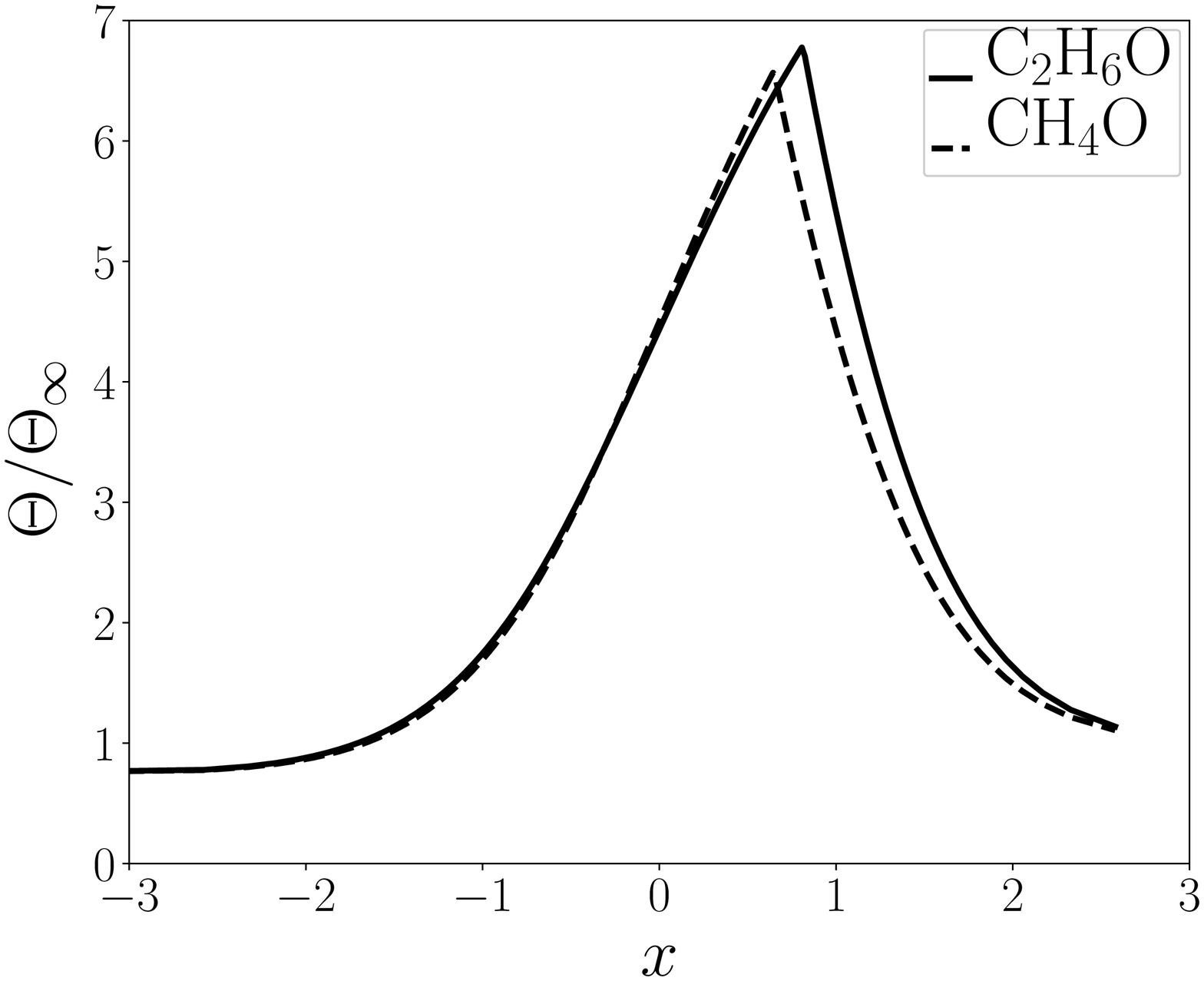}
}
\resizebox*{\textwidth}{!}{
\includegraphics{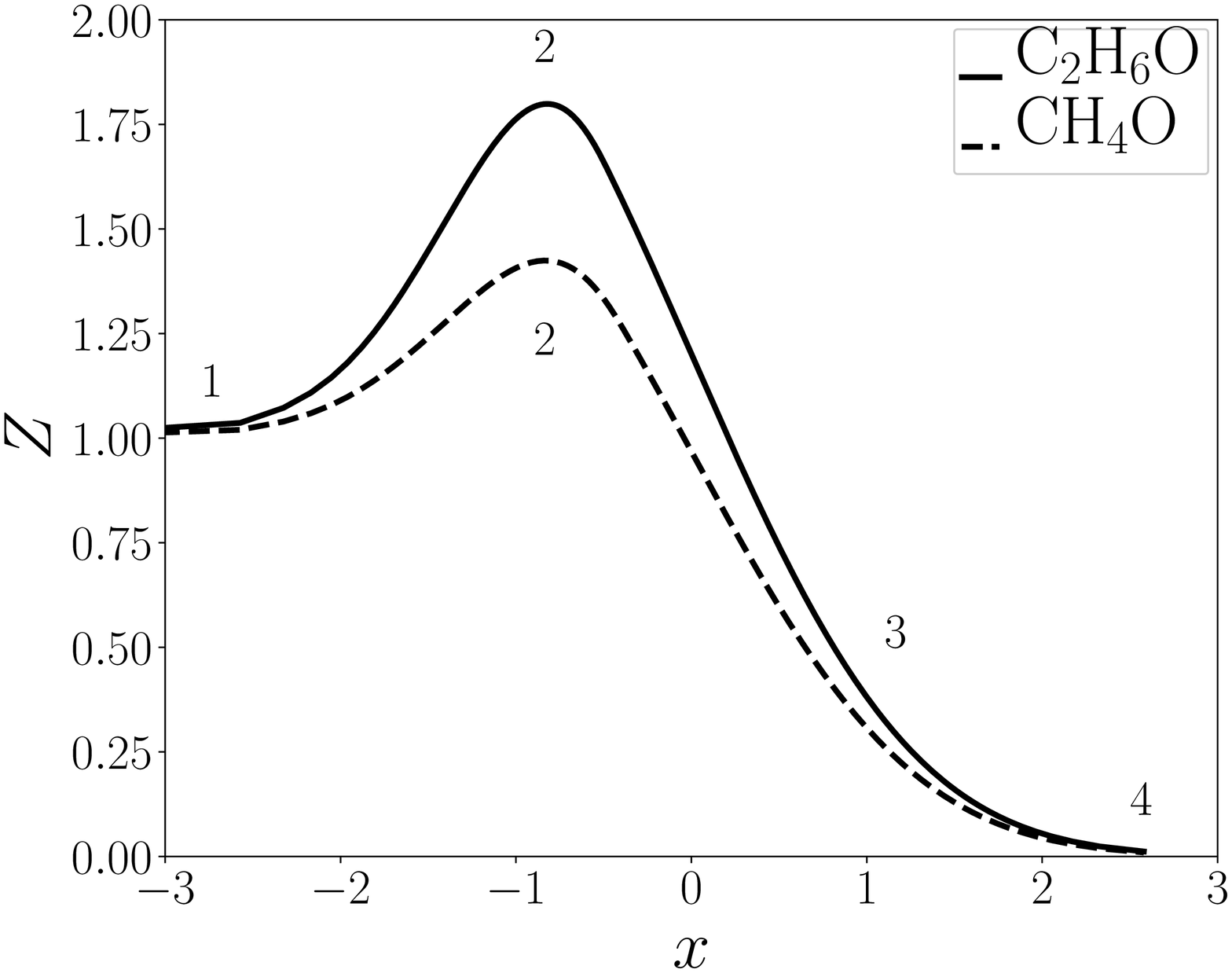}
\includegraphics{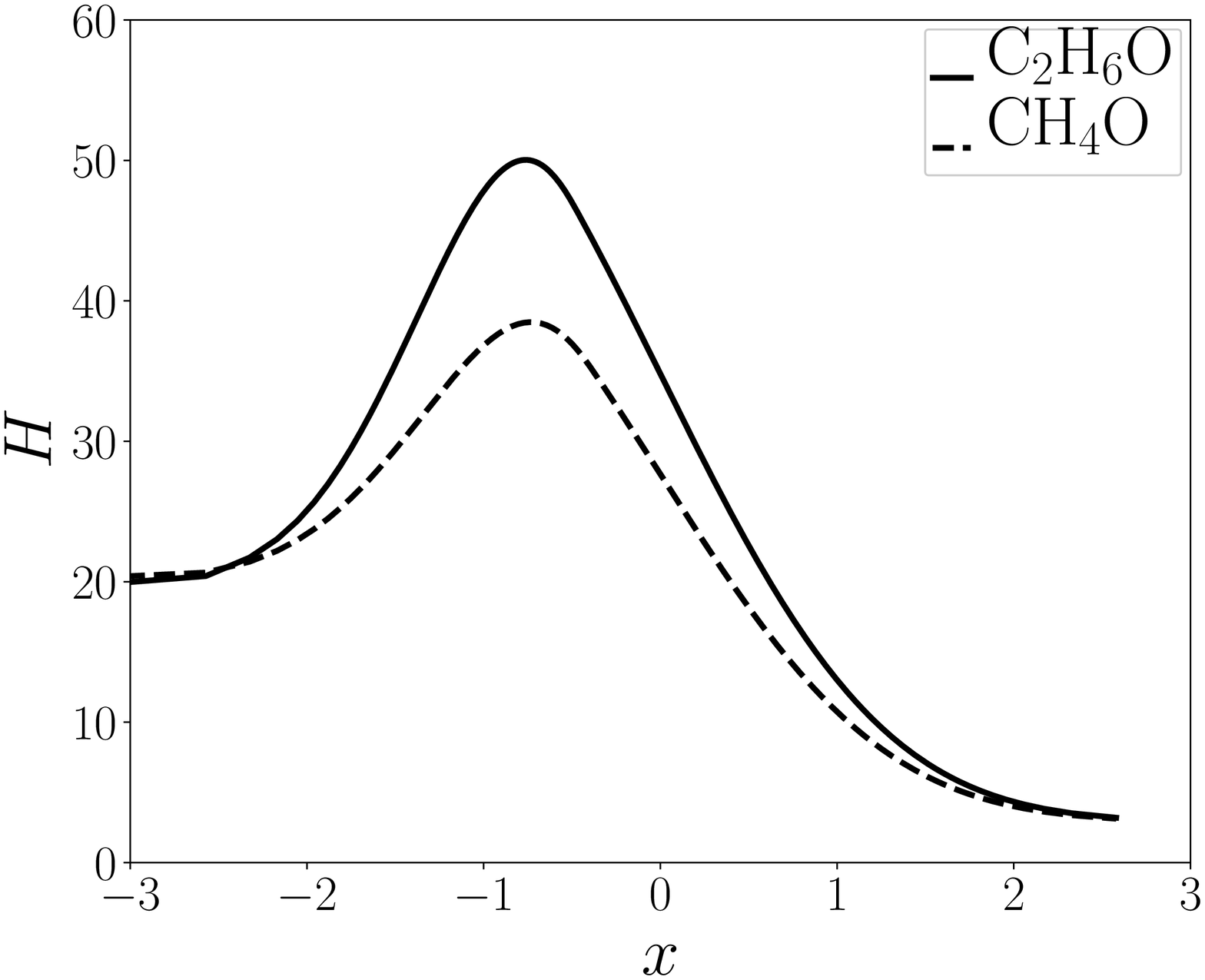}
}
\caption{Mass fractions ($Y_F$, $Y_O$), temperature ($\Theta/\Theta_\infty$), mixture fraction ($Z$) and excess of enthalpy ($H$) profiles in the physical space $x$ . The peak observed in the profiles of $Y_F$, $Y_O$, $Z$ and $H$ is due to the vaporisation source. The flame position around $x_f = 0.8$ separates the fuel region at left ($Y_O = 0$)  from the oxidiser side at right ($Y_F = 0$). }\label{fig03}
\end{center}
\end{figure}

All results, from here on, were obtained in $Z_C-$space and subsequently mapped to $x-$space for clarity, unless specified otherwise.  
Fig.~\ref{fig03} shows profiles of $Y_{F}$, $Y_{O}$, $\Theta$, $Z$ and $H$.
The flame position is characterised by $Y_O = Y_{F} = 0$ which expectedly coincides with the location where the temperature is highest.
The peak observed in the mass fractions is due to the vaporisation of the droplets since it is a source of mass. 
A few things are worth mentioning from these plots:
(i) C$_2$H$_6$O has a lower latent heat of vaporisation than CH$_4$O, as a result its mass fraction reaches higher values;
(ii) Additionally, C$_2$H$_6$O has a higher heat of combustion than CH$_4$O which leads to a higher flame temperature than for CH$_4$O;
(iii) The flame achieves stoichiometric conditions further into the oxidant side for C$_2$H$_6$O than for CH$_4$O, this is due to the higher fuel content for the former.

\begin{figure}[h!]
\begin{center}
\resizebox*{\textwidth}{!}{
{\includegraphics{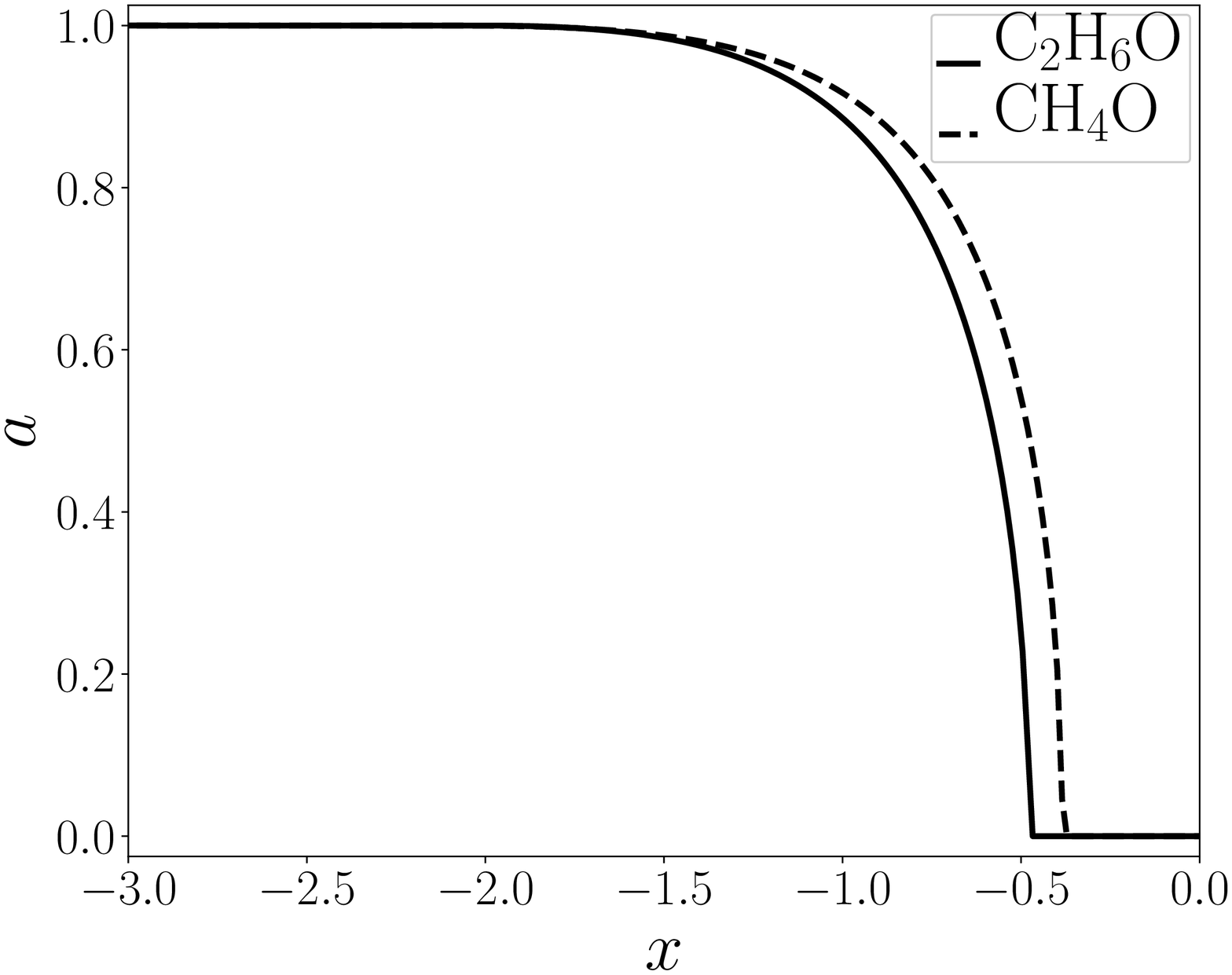}}
{\includegraphics{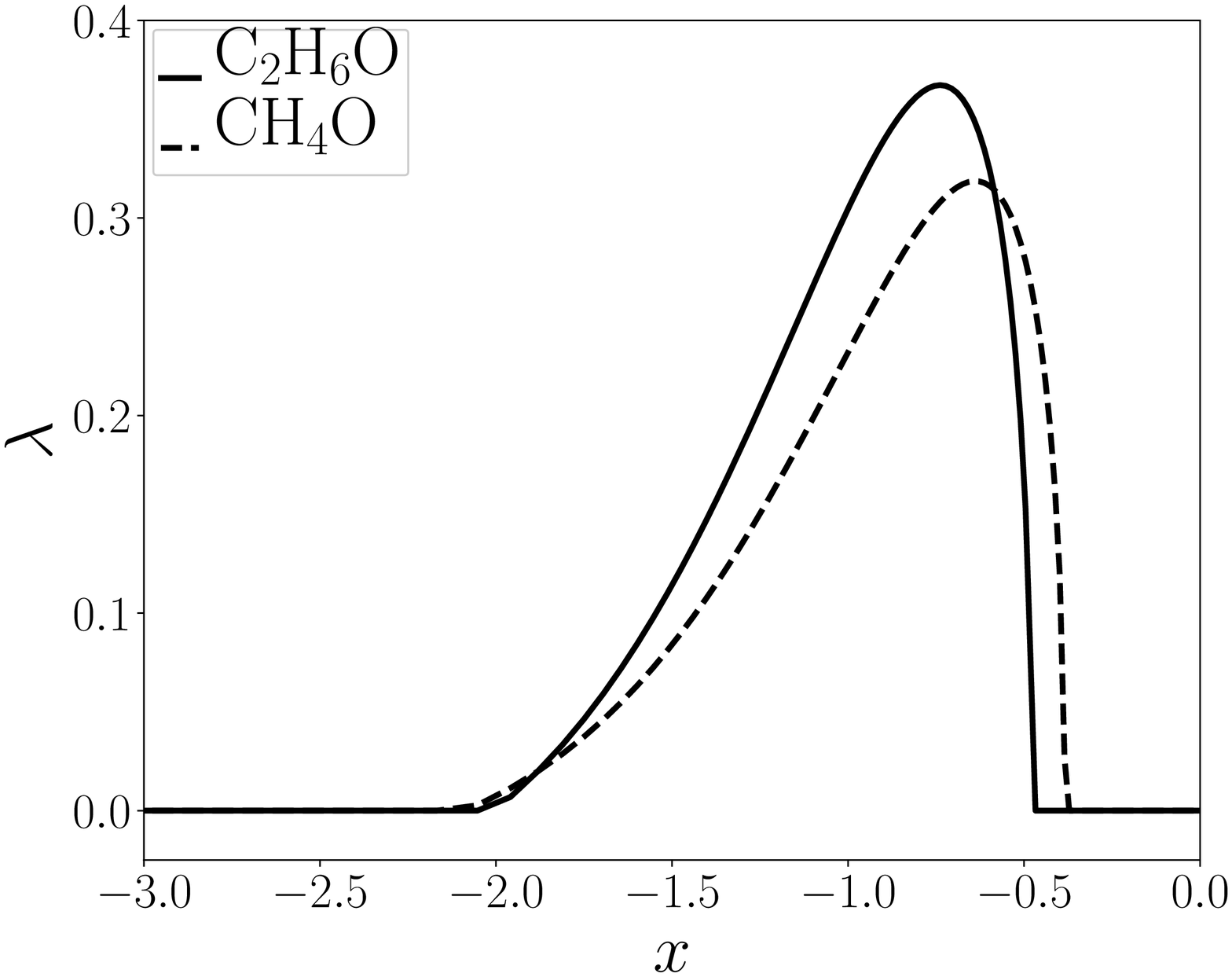}}
}
\hspace{0.1\textwidth} (a)\hspace{0.45\textwidth} (b) 
\caption{(a) Droplet radius $a$ and (b) vaporisation rate $\lambda$, which is non-zero only in the region where droplets vaporise.}\label{fig04}
\end{center}
\end{figure}

The droplet radius and evaporation rate spatial distributions are presented in Fig.~\ref{fig04}, as in \cite{Maionchi2013}.
The droplet radius is initially constant, and decreases as it approaches the flame  until the droplet is fully vaporised (see Fig.~\ref{fig04}~(a)).
The vaporisation extends further for CH$_4$O than for C$_2$H$_6$O (i.e. $a=0$ at $x= -0.38$ and $x= -0.47$, respectively.)
Fig.~\ref{fig04}~(b), is in line with the result obtained for the droplet radius, since the evaporation rate, $\lambda$, is non-zero only in the region where the droplets are present.

\begin{figure}[]
\begin{center}
\resizebox*{\textwidth}{!}{
\includegraphics{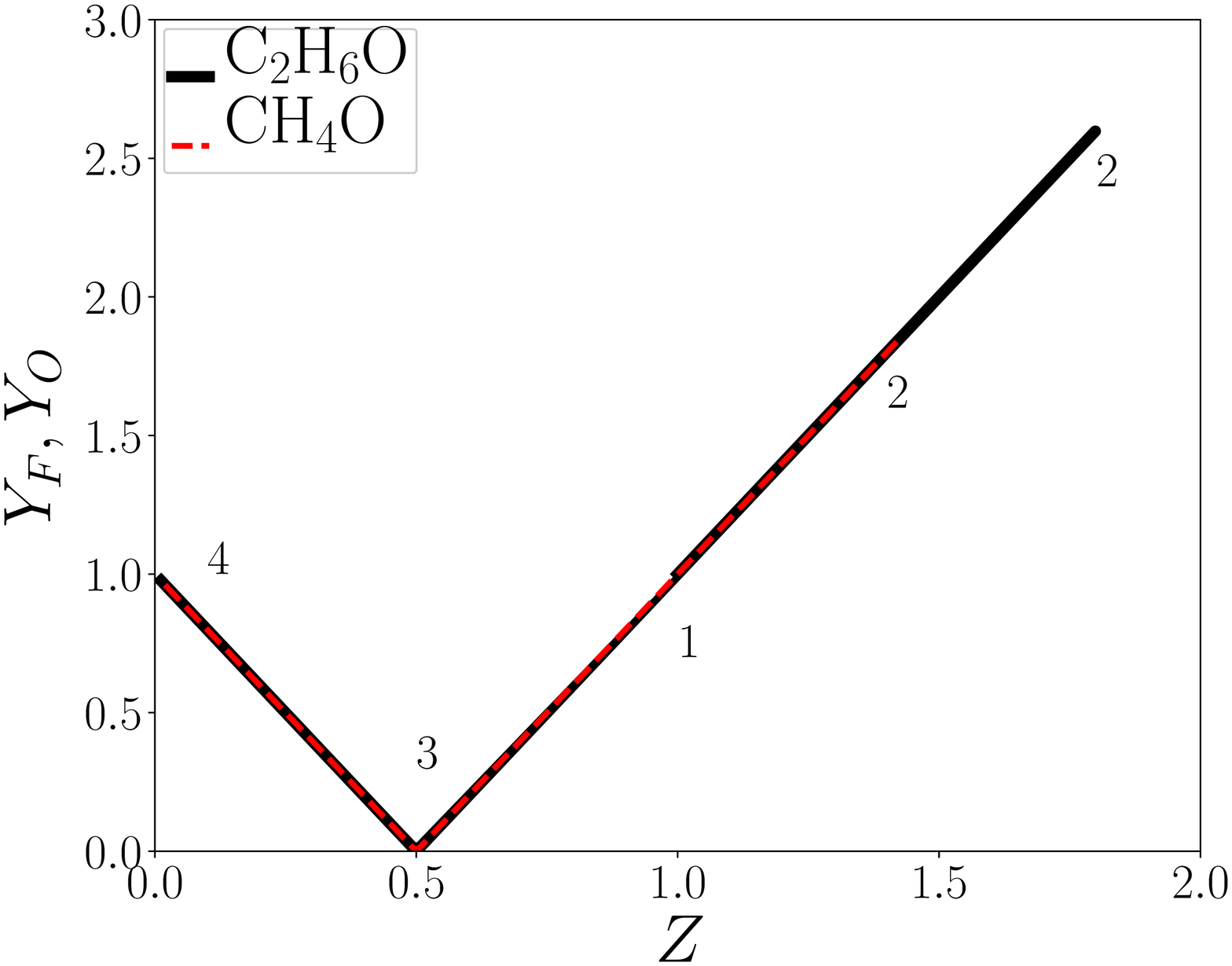}
\includegraphics{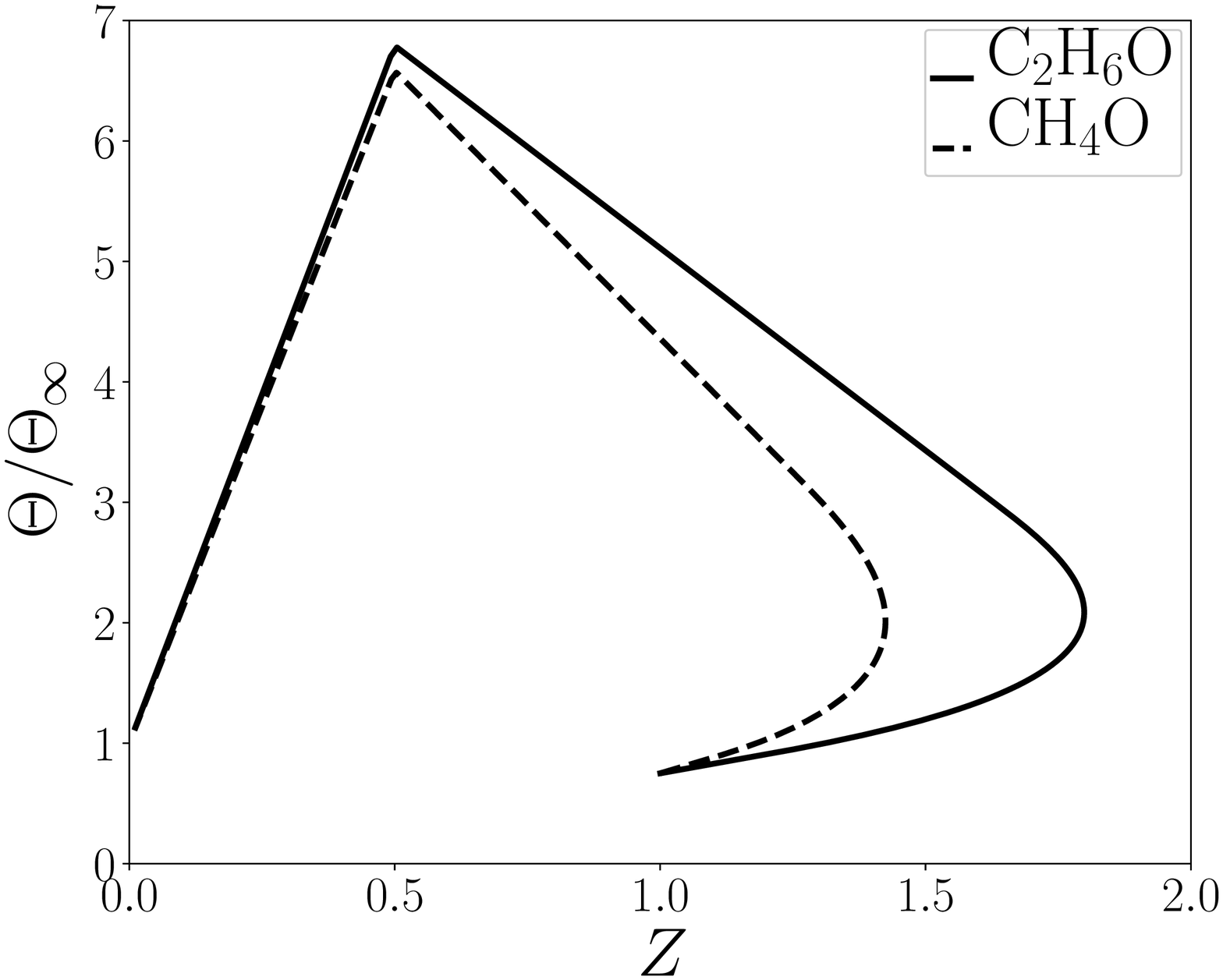}
}
\resizebox*{\textwidth}{!}{
\includegraphics{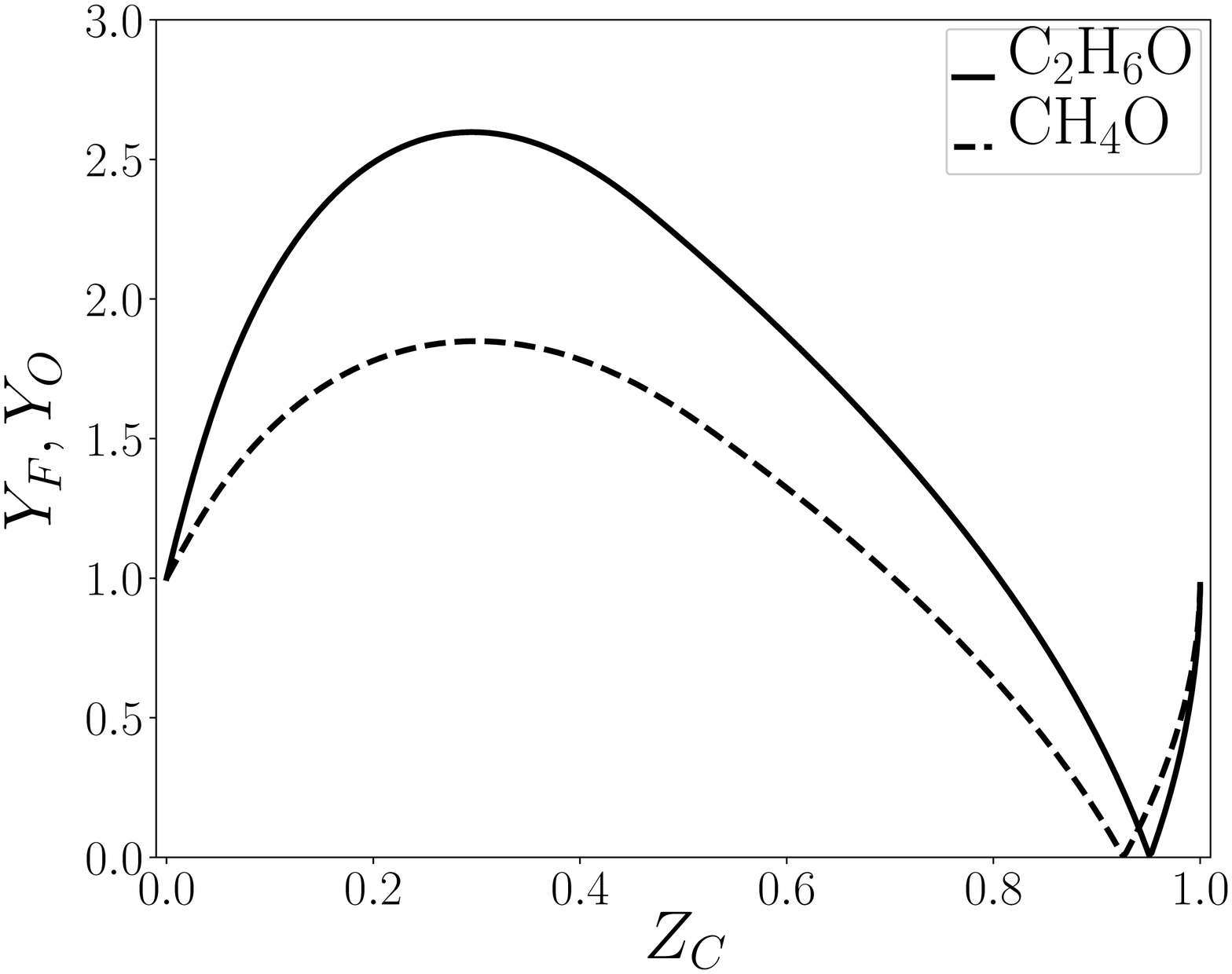}
\includegraphics{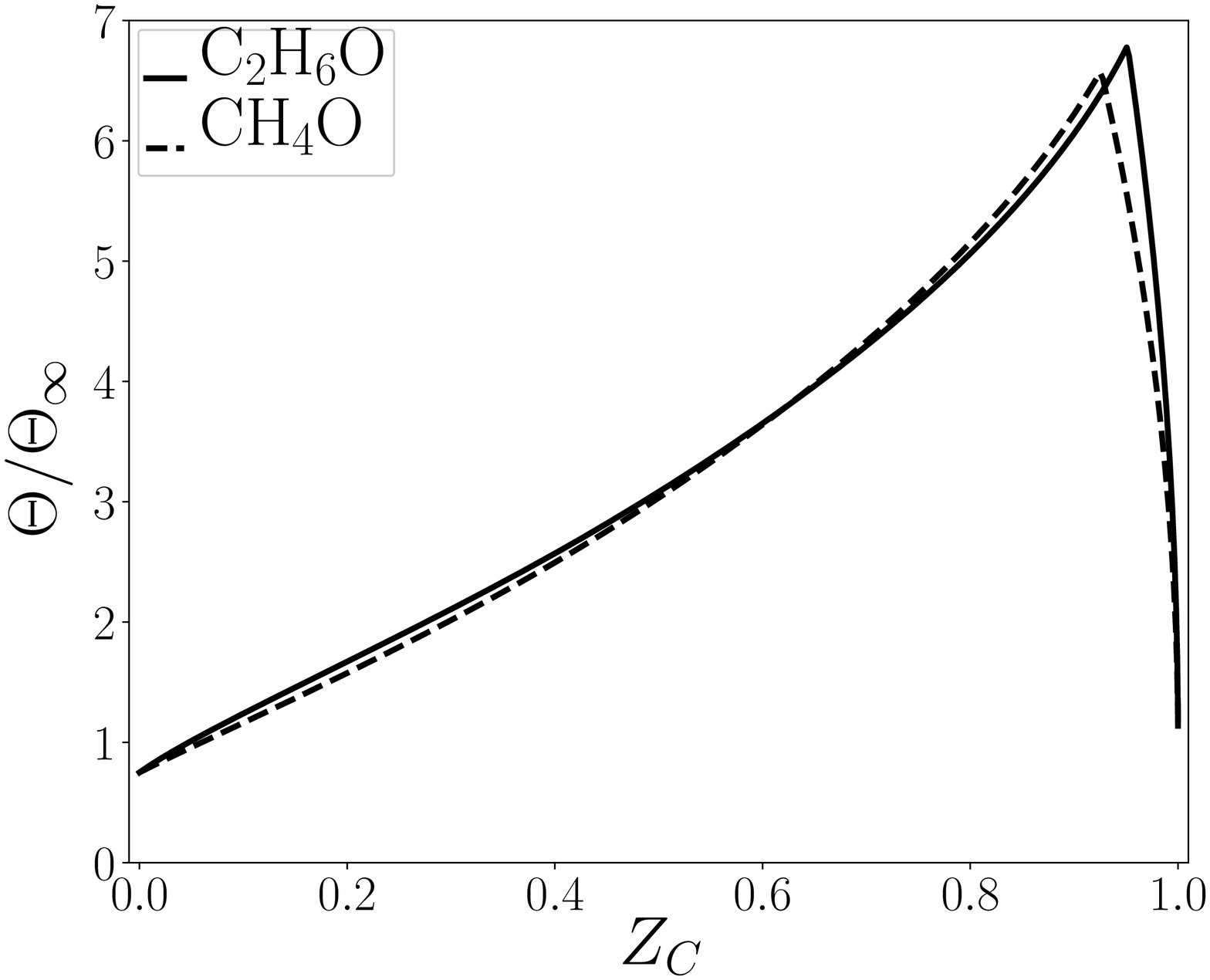}
}
\resizebox*{\textwidth}{!}{
\includegraphics{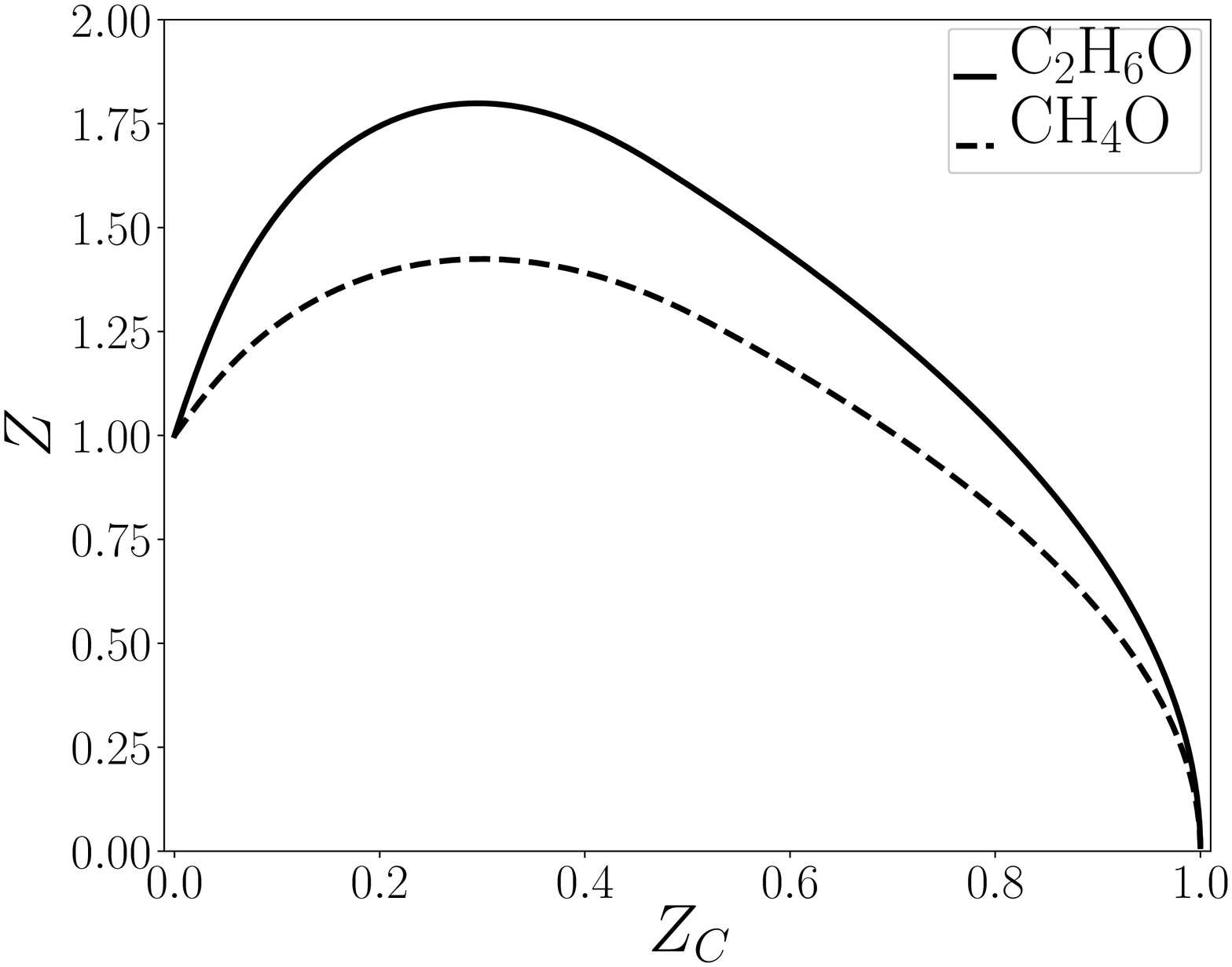}
\includegraphics{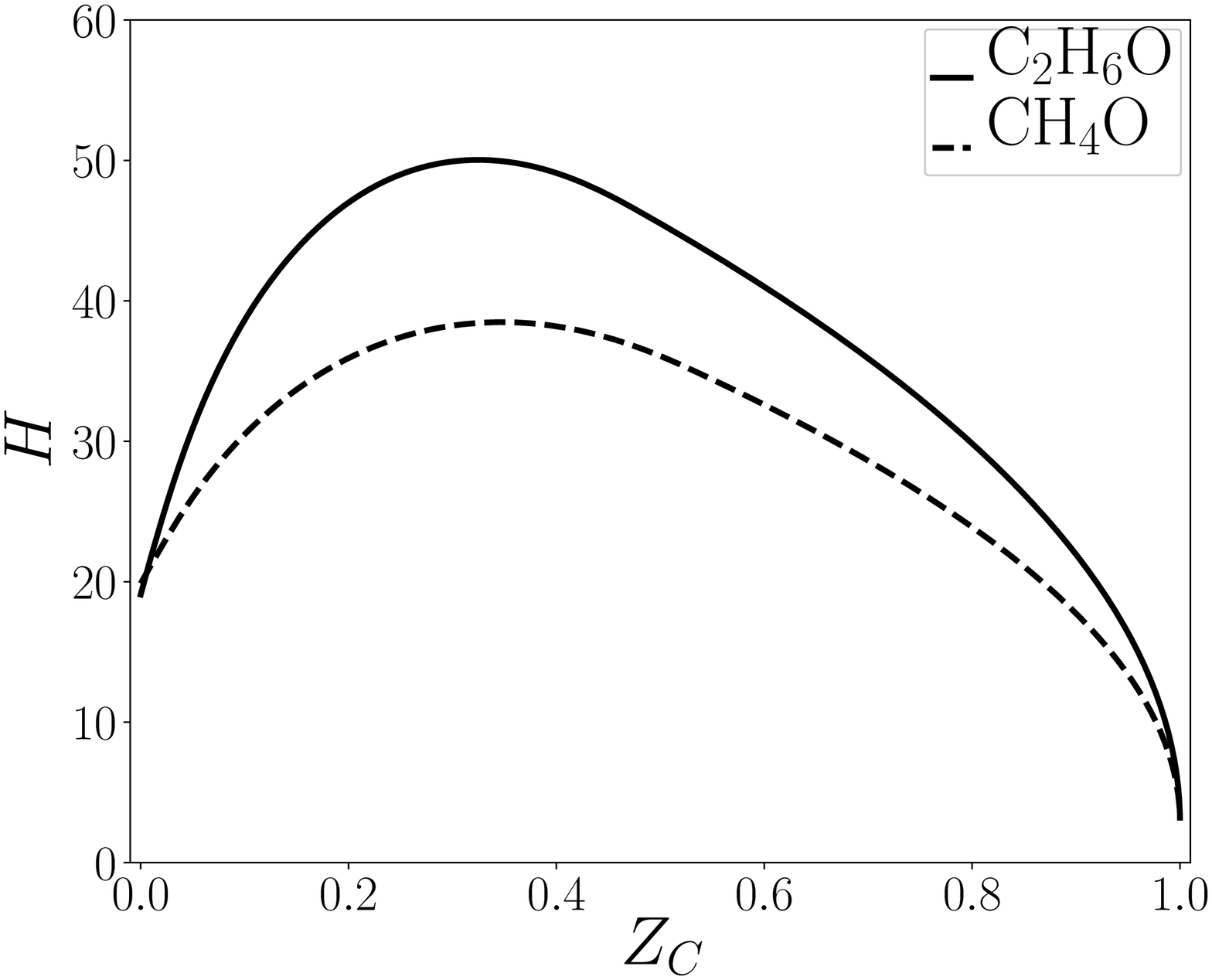}
}
\caption{mass fractions ($Y_F$ , $Y_O$) and temperature profiles ($\Theta/\Theta_\infty$) as a function of the mixture fraction $Z$ (Top). For $Z > 1$, which corresponds to the peak of the vaporisation zone in Fig.~\ref{fig03}, $Y_F$ nor $\Theta$ are uniquely defined, showing the non-monotonic behaviour in the $Z$-space description of the problem. Annotations: 1: beginning of vaporisation in the fuel side ($Z=1$), 2: maximum in vaporisation rate (maxima in $Y_F$ and $Z$), 3: flame position ($Y_F = Y_O = 0$), 4: incoming flow oxidant side ($Z=0$);
Mass fractions ($Y_F$, $Y_O$) and temperature profiles ($\Theta/\Theta_{\infty}$) (Middle), mixture fraction ($Z$) and excess of enthalpy ($H$) (Bottom) in the $Z_C$-space for $Le_F = Le_O = 1$. Note that the temperature and the mass fraction are uniquely defined in this space, as opposed to when they are defined  in the $Z$-space.}
\label{fig04p}
\end{center}
\end{figure}

To highlight the value of the cumulative mixture fraction, $Z_C$, proposed in this work, Fig.~\ref{fig04p} shows a comparison of the profiles of mass fractions and temperature in $Z$ and $Z_C$-space.
The profiles in $Z$-space for each fuel have the same linear dependence as that given from its definition at stoichiometric conditions $Z = (1 + Y_F - Y_O)/2$, valid for unity Lewis numbers.
The maximum value of $Z$ attained for each fuel differs (see red dashed line and solid black line), as these values are a function of the vaporisation rate.


Close inspection of Fig.~\ref{fig04p}, shows that for $Z>1$, the mass fractions and temperature profiles are multivalued.
Note that in contrast with pure gas flow whose maximum value is bounded at $Z=1$, here $Z$ extends beyond unity.
The non-monotonicity for the $Y_F$ variable occurs along the same straight line and thus, it is not visible at this scale in Fig.~\ref{fig04p}.
The numbered annotations in Fig.~\ref{fig04p} correspond to the path taken by the fuel stream in the $Z-$space, from injection until it reaches the flame.
At position $1$ ($Z=1$) the fuel droplets are injected in the gaseous fuel stream.
The vaporisation of droplets increases the value of $Y_F$ to a maximum, different for each fuel (see position $2$).
From point $2$ to $3$, consumption of fuel by the flame decreases the value of $Y_F$, from its maximum at point $2$ to zero at point $3$.
From the description above, it is evident that $Z$ is not an adequate function that guarantees that both $Y_F$ and $\Theta$ be uniquely defined, as the trajectories taken by the fuel mass fraction and temperature profiles (path $1-2-3$) are non-monotonous in $Z-$space.
For clarity, path $1-2-3$ is also shown in Fig.~\ref{fig03} in physical space, $x$.
Finally, it is worth reiterating that remapping the solution to physical space, $x$, would not be possible due to the multivalued nature of the formulation in $Z$-space. 
%

%
%
%

Figure~\ref{fig04p} also shows the profiles of mass fractions and temperature as a function of $Z_C$; profiles of $Z$ and $H$ are also shown for completeness. 
In this space, $Z_C = 0$ corresponds to the fuel stream whereas $Z_C = 1$ represents the oxidant stream.
The flame position ($Y_O = Y_F = 0$) can be identified by searching for the value of $Z_C$ where the normalised temperature, $\Theta/\Theta_\infty$, reaches its maximum value.
These results show clearly that the variables are single-valued in the $Z_C$-space and that $Z_C$ is a useful space for the spray-flamelet description.

It must be emphasised that the profiles do not show over- or under-estimated values of temperature, regardless of the droplet radius or strain rate, which is accounted for through $M$.
This is an advantage over \citep{Franzelli2015} where the temperature is overestimated for the highest values of the strain rate due to the assumed closure relation for $\chi$. 
In the present work, this is not the case because $\chi$ can be directly evaluated from the formulation, \textcolor{blue}{as seen in Eq. (\ref{eq21a})}.

\begin{figure}[h!]
\begin{center}
\resizebox*{\textwidth}{!}{
\includegraphics{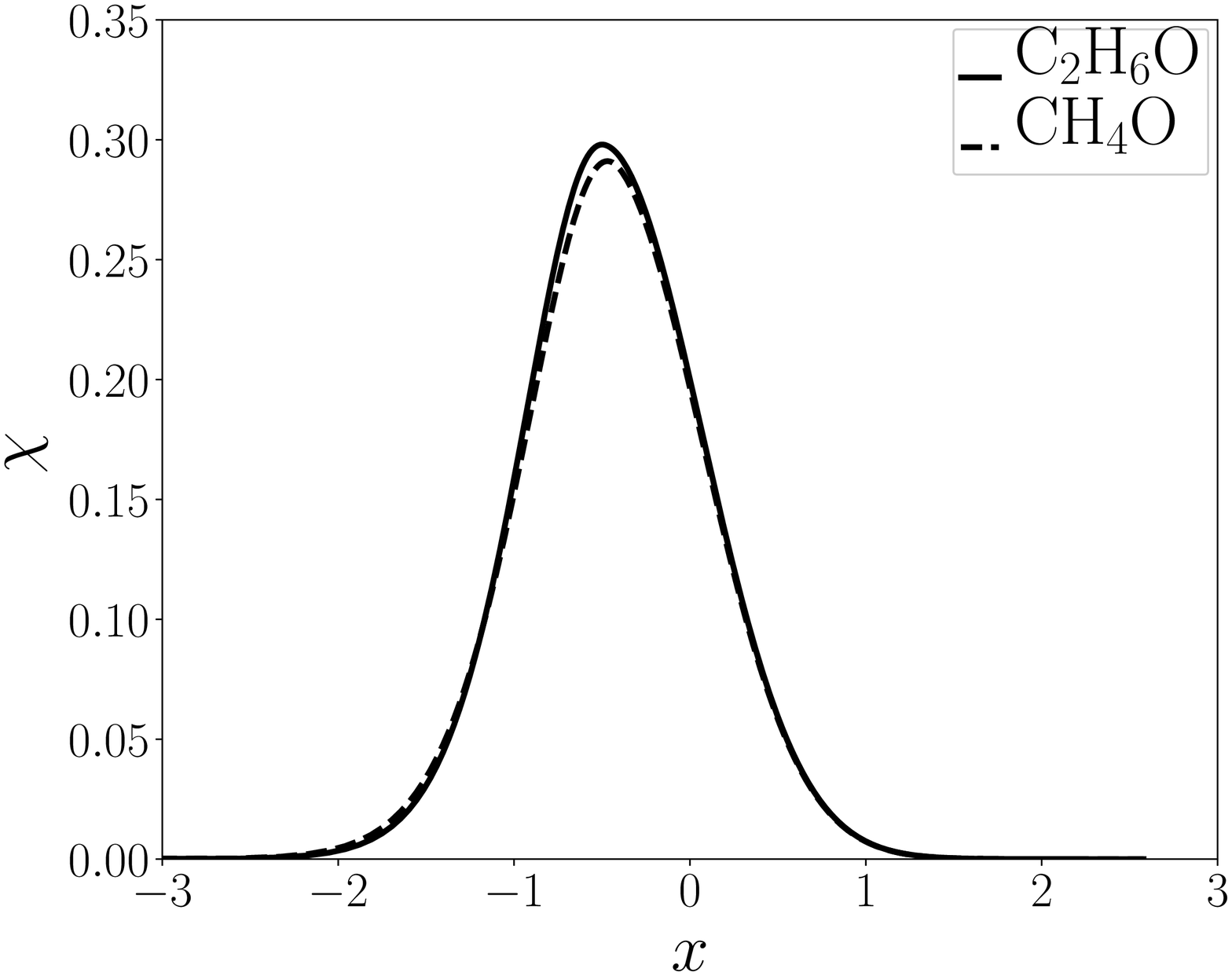}
\includegraphics{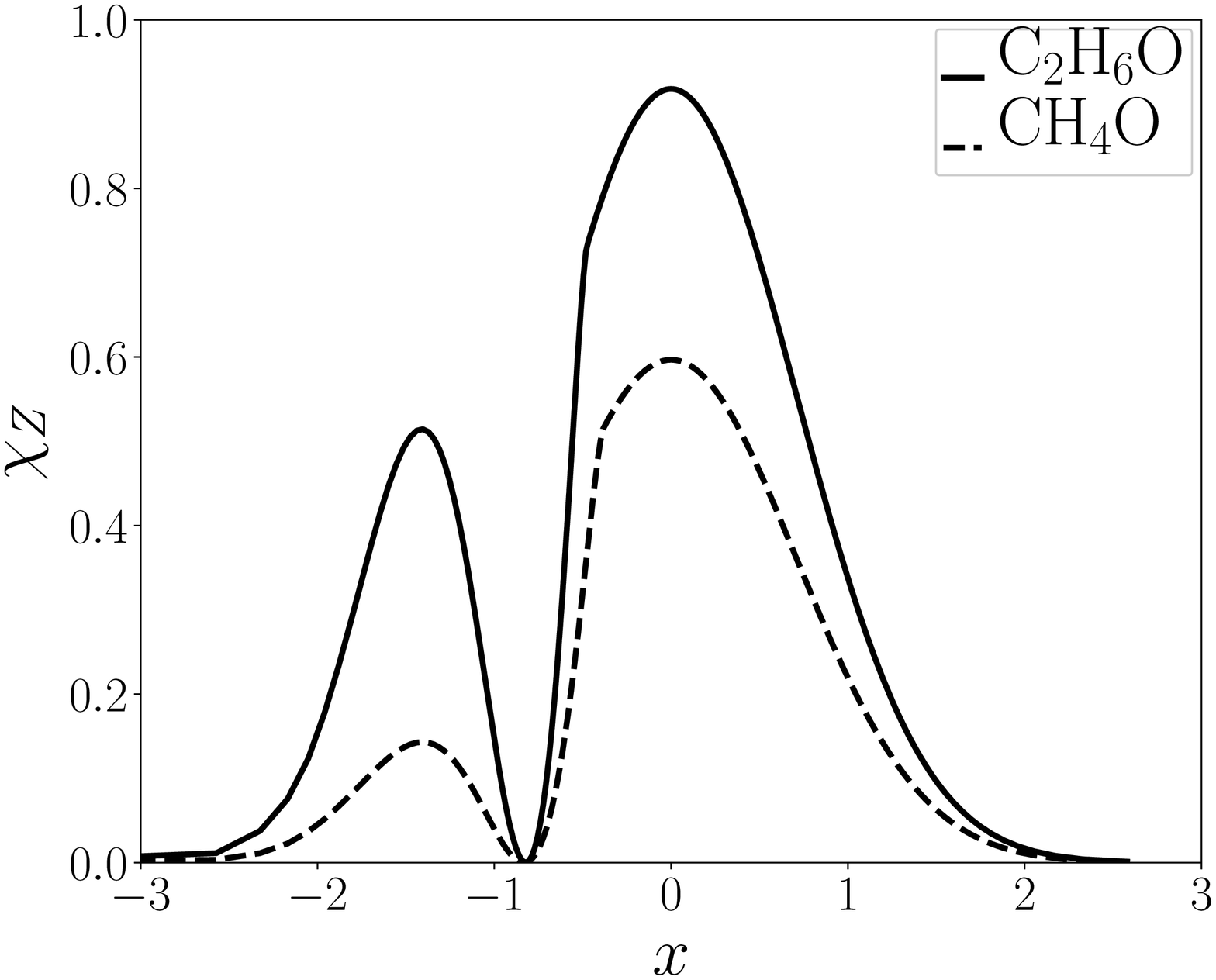}
}
\hspace{0.1\textwidth} (a)\hspace{0.45\textwidth} (b) 
\caption{The scalar dissipation rate for sprays (a) $\chi = (dZ_C/dx)^2$ and (b) $\chi_Z = (dZ/dx)^2$ using $Z_C$ and $Z$ respectively as the generic variable. $\chi_Z $ is the definition typically considered for purely gaseous flow.}\label{fig06}
\end{center}
\end{figure}

Finally, the comparison of the scalar dissipation rate, $\chi = (dZ_C/dx)^2$ , and that obtained using $Z$ as the generic variable, $\chi_Z = (dZ/dx)^2$ \cite{Peters1984}, are shown in Fig.~\ref{fig06}. 
Since in the present case the flow is aligned with the physical axis, such that $d\eta/dx = 1$, the closed-form expression for the scalar dissipation rate $\chi$ in space $Z_C$ is obtained from Eq. (\ref{eq21a}) as
\begin{equation}
 \chi = 2D \left(\frac{1}{Z_C^T}\frac{e^{-\eta^2/2}}{\sqrt{2\pi}}Z(\eta)\right)^2.
 \label{eq.chi}
\end{equation}
In agreement with Eq. (\ref{eq.chi}), Fig.~\ref{fig06} shows that $\chi$ increases smoothly, reaches a maximum value and then decreases almost symmetrically. 
In contrast, $\chi_Z$ exhibits a more complex behavior (see Fig.~\ref{fig06}~(b) ),
these shapes can be understood by inspecting Eq. (\ref{eq.chi}) and recalling the results presented in Fig.~\ref{fig02}: the maximum value of $Z$ explains why there is a region where $\chi_Z$ goes to zero; the profile of $Z_C$ as a function of $x$ has an inflection point responsible for the maximum value observed in profiles shown for $\chi$.
%


\subsection{Non-unity $Le_F$ and $Le_O$}
\label{subsec:Le_effects}

\begin{table}[h!]
\centering
\caption{Lewis numbers.}\label{tab02}
\begin{tabular}{ccc}
\hline
    O$_2$ & CH$_4$O  & C$_2$H$_6$O\\
\hline
    1.06 & 1.25 & 1.73 \\
\hline
\end{tabular}
\end{table}

\begin{figure}[ht!]
\begin{center}
\resizebox*{\textwidth}{!}{
\includegraphics{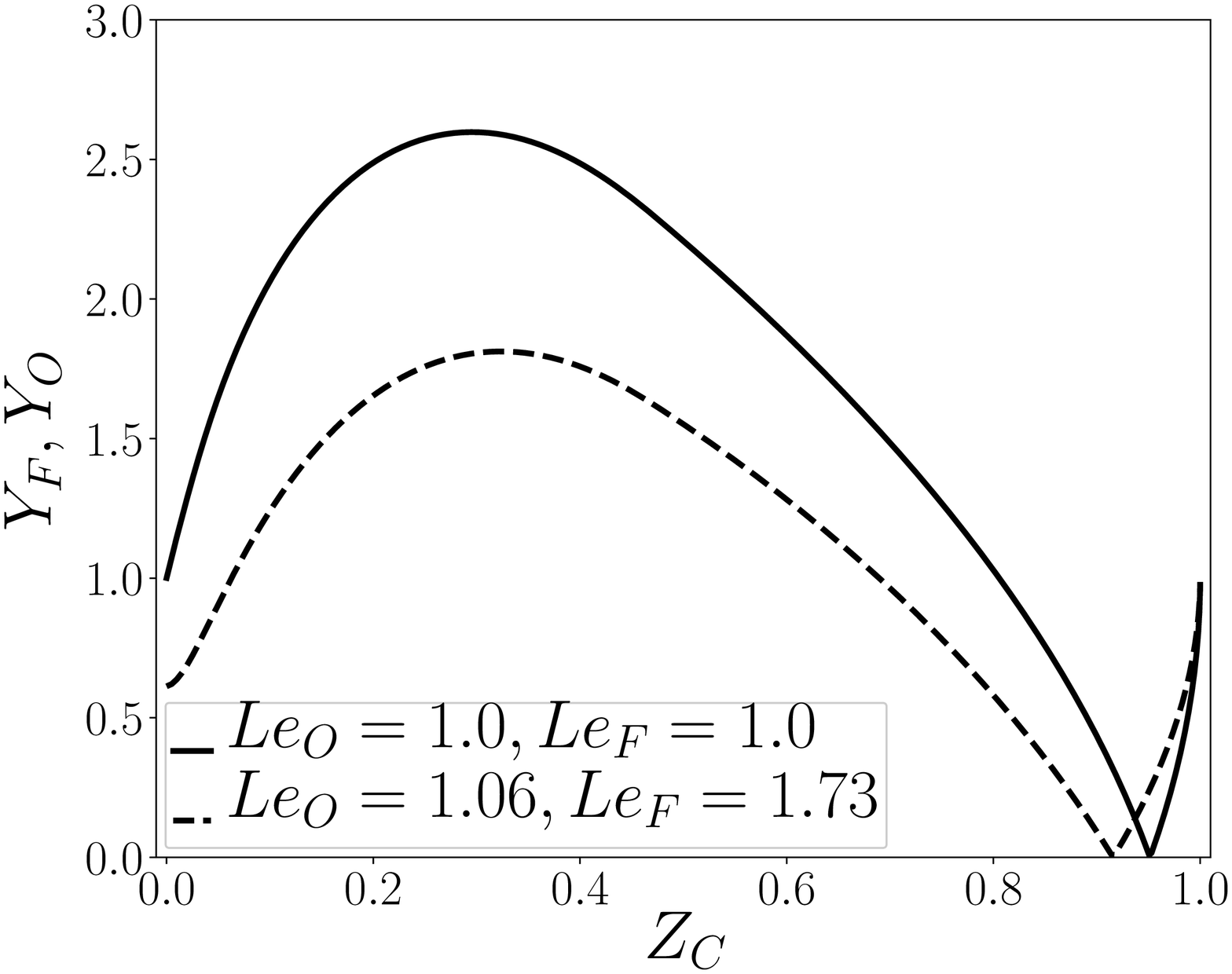}
\includegraphics{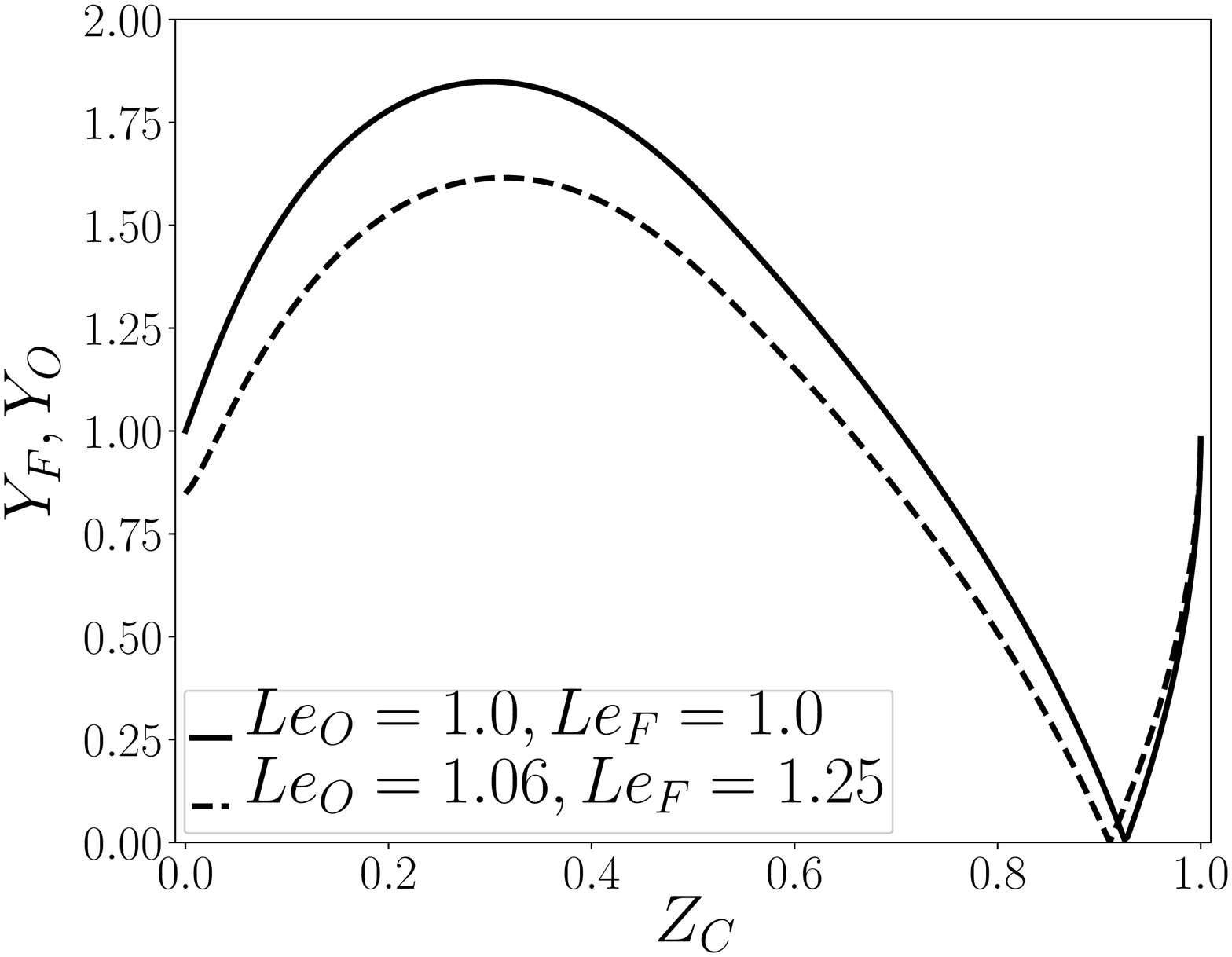}
}
\resizebox*{\textwidth}{!}{
\includegraphics{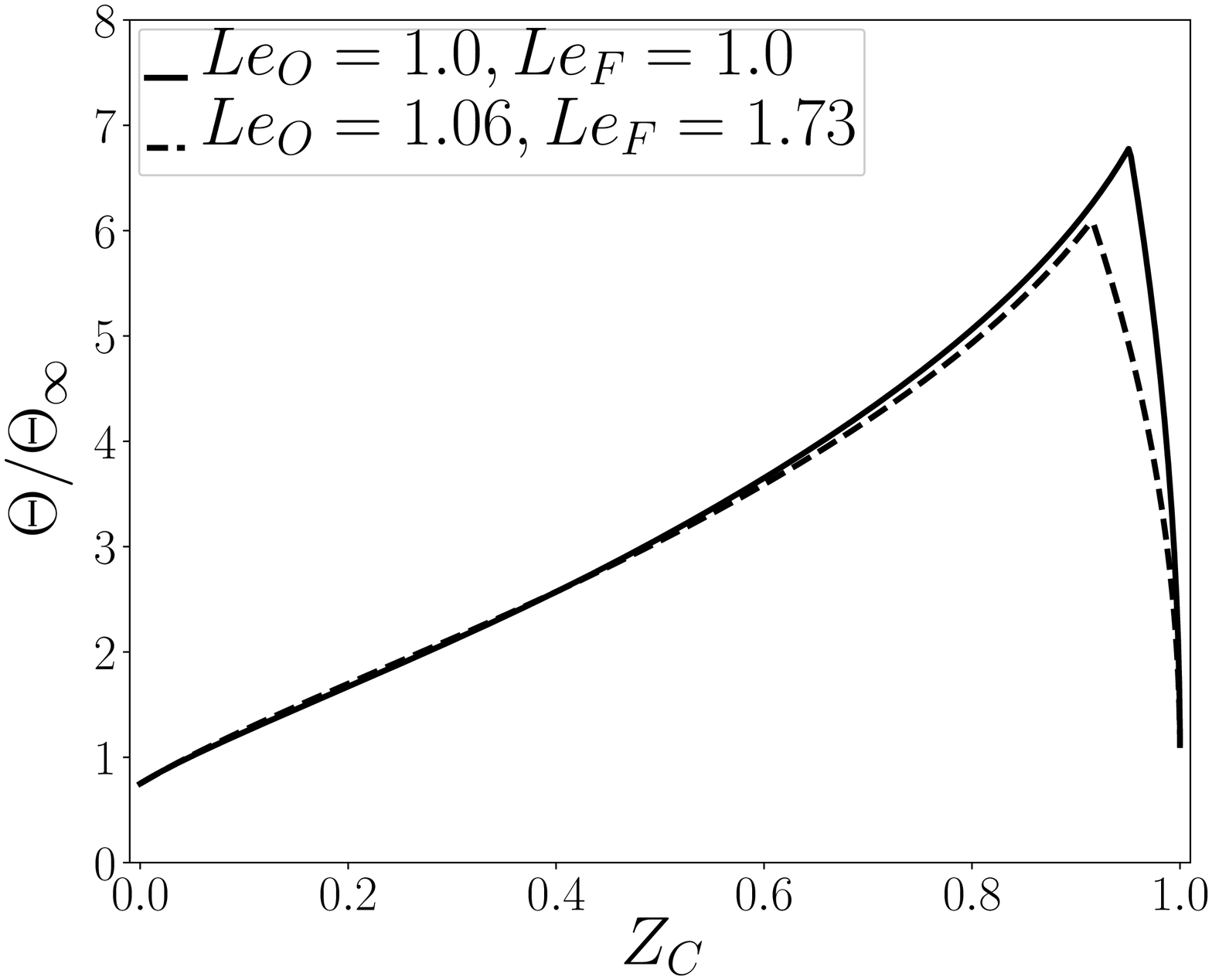}
\includegraphics{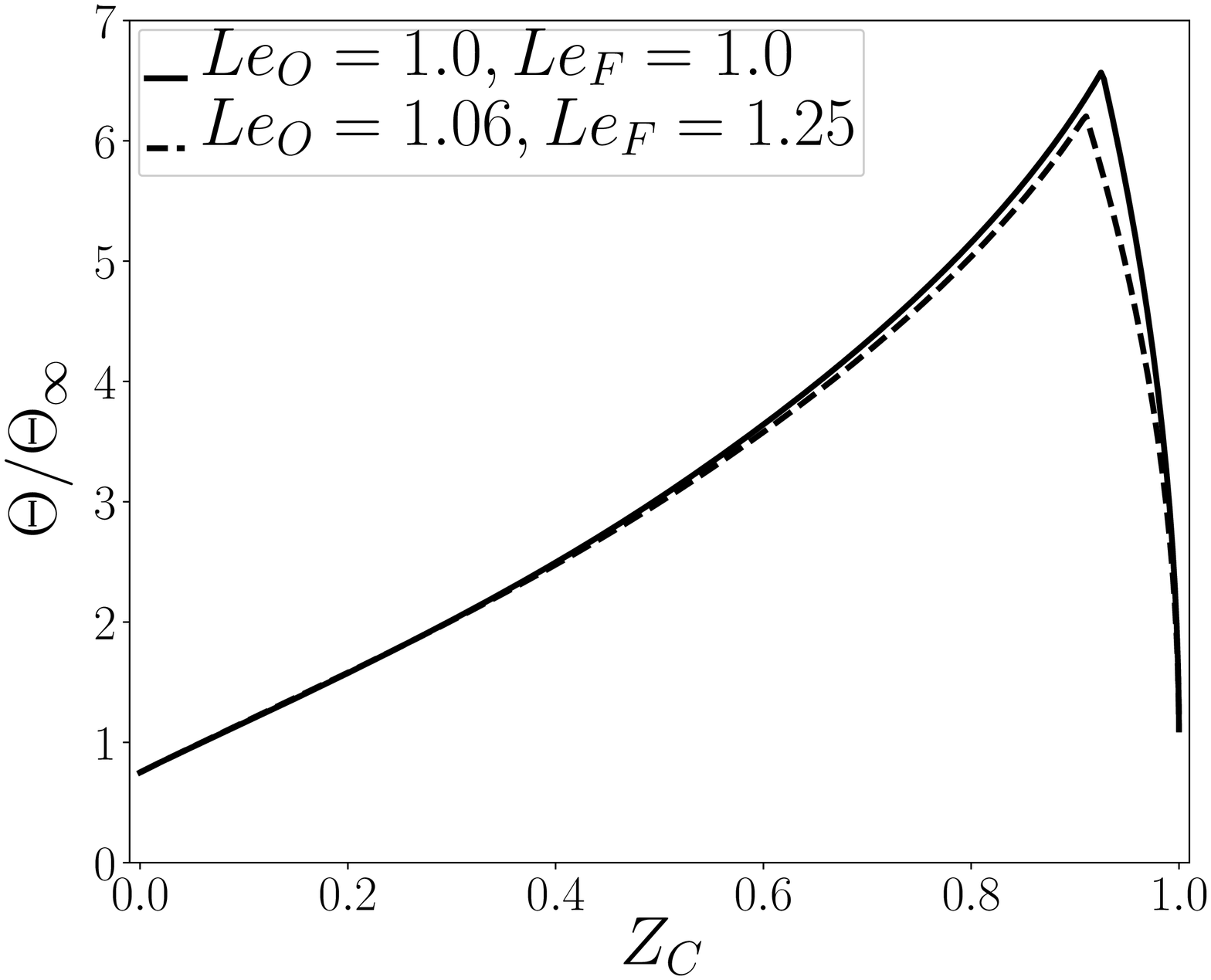}
}
\resizebox*{\textwidth}{!}{
\includegraphics{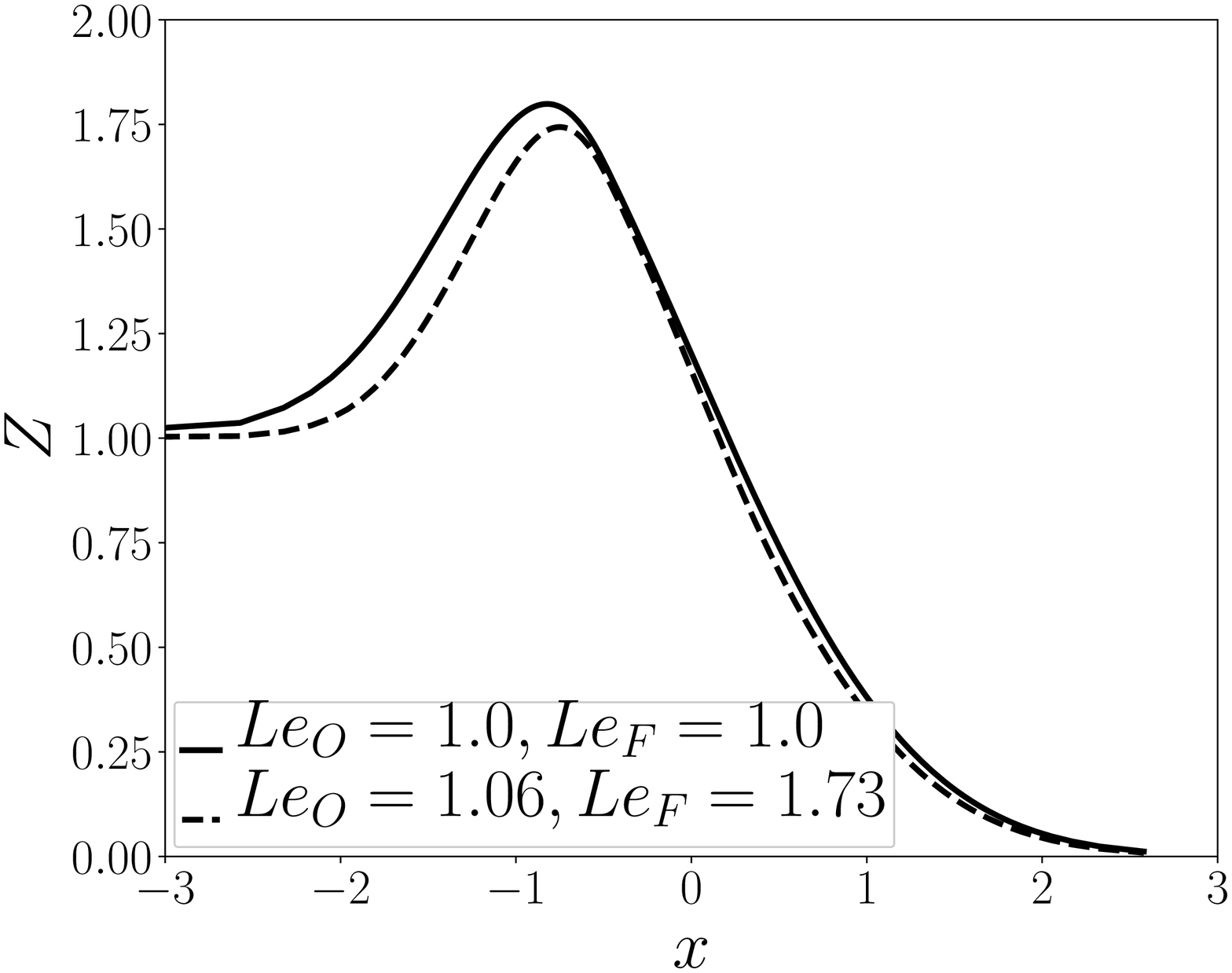}
\includegraphics{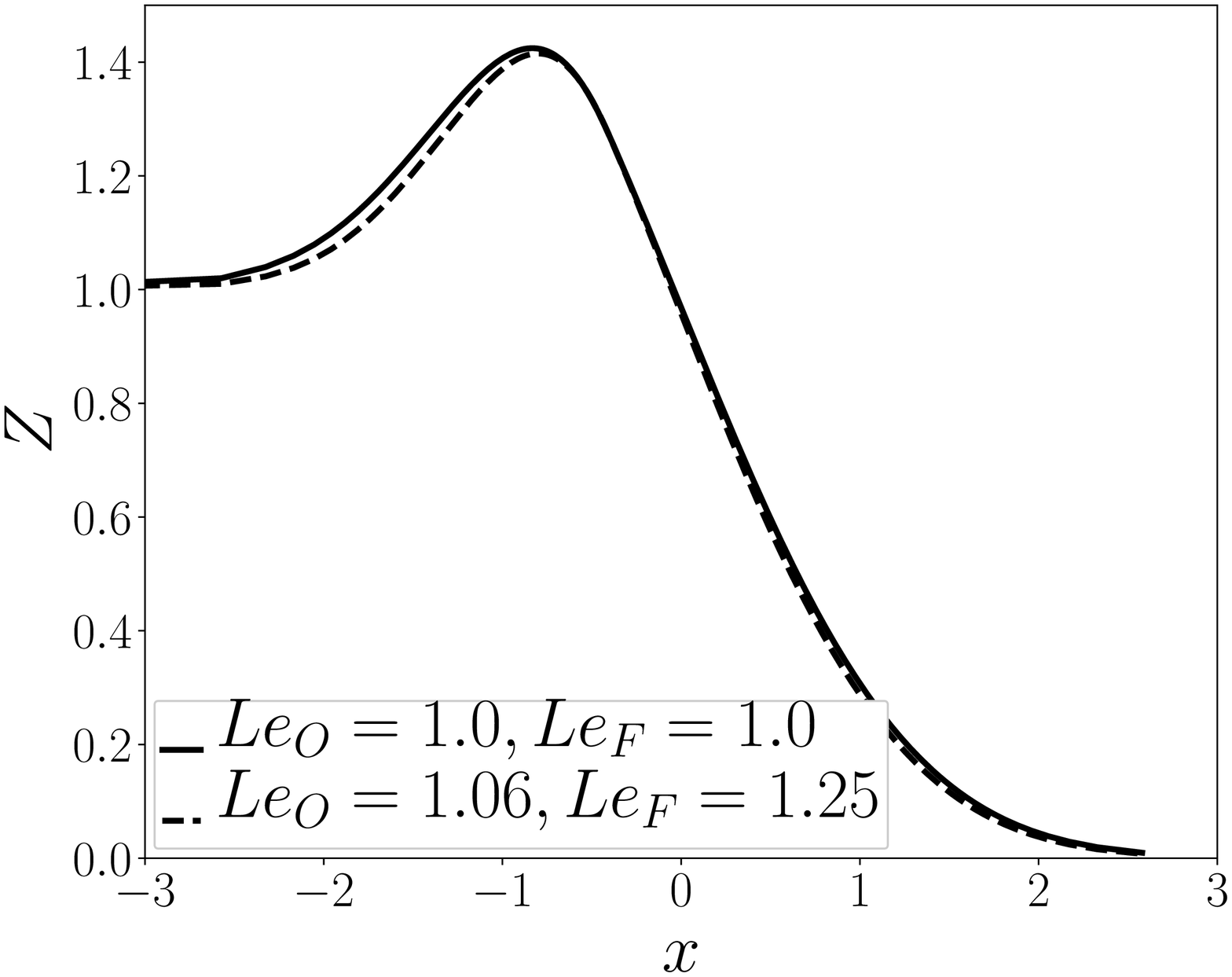}
}
\hspace{0.1\textwidth} (a)\hspace{0.45\textwidth} (b) 
\caption{Comparison between the results with unitary and correct values of the Lewis numbers for (a) C$_2$H$_6$O (ethanol) and (b) CH$_4$O (methanol)}.\label{fig07}
\end{center}
\end{figure}

In this section we briefly assess the effect of non-unity $Le$ numbers in the spray-flamelet structure.
The $Le$ numbers used in the simulations are presented in Table~\ref{tab02}. These were determined using Cantera~\cite{goodwin2009cantera}, and CaltechMech~\cite{narayanaswamy2010consistent}. %
Profiles of $Y_{F}$, $Y_{O}$, $\Theta$, and $Z$ in $Z_C$-space, are shown in Fig.~\ref{fig07} for both fuels C$_2$H$_6$O and CH$_4$O.
The profiles show quantitative differences when compared with the unity $Le$ number results (solid lines); these differences are more evident for C$_2$H$_6$O, which has a higher $Le_F$ than CH$_4$O. 
While the differences in $\Theta$ and $Z$ do not seem to be as pronounced for both fuels, the mass fractions show a more significant variation. Also, the flame position sensitivity increases with increasing $Le$. 

\begin{figure}[h!]
\begin{center}
\resizebox*{\textwidth}{!}{
\includegraphics{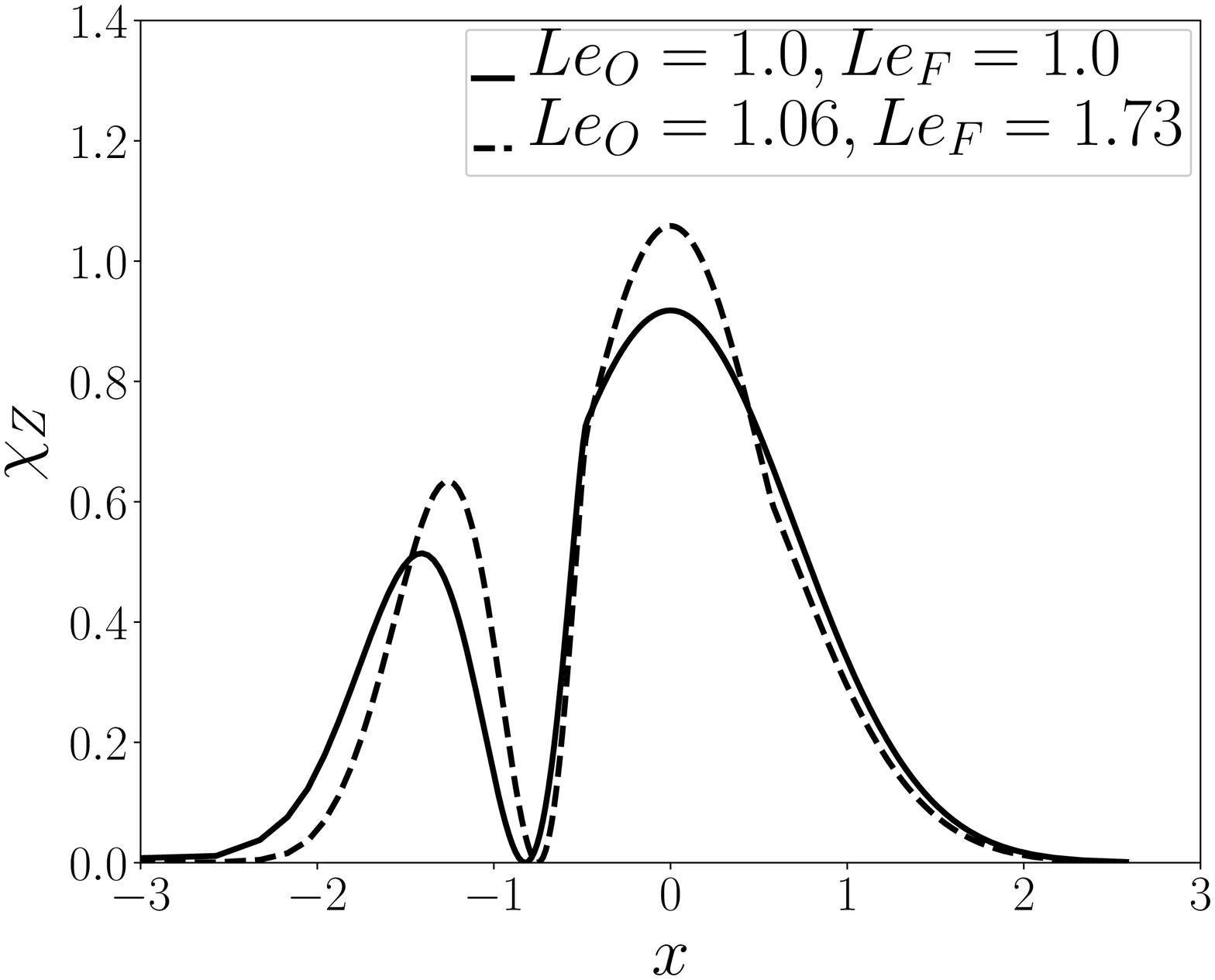}
\includegraphics{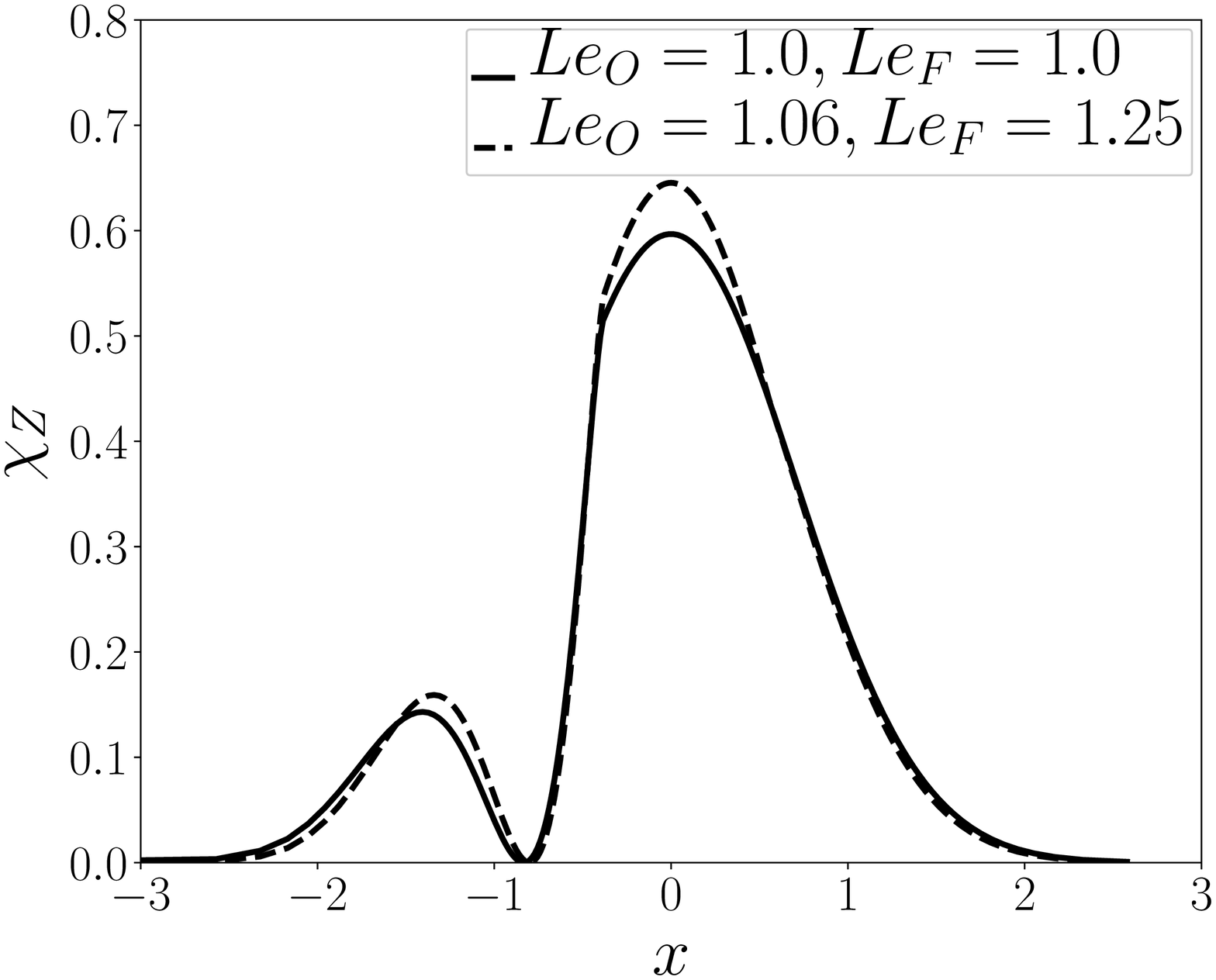}
}
\resizebox*{\textwidth}{!}{
\includegraphics{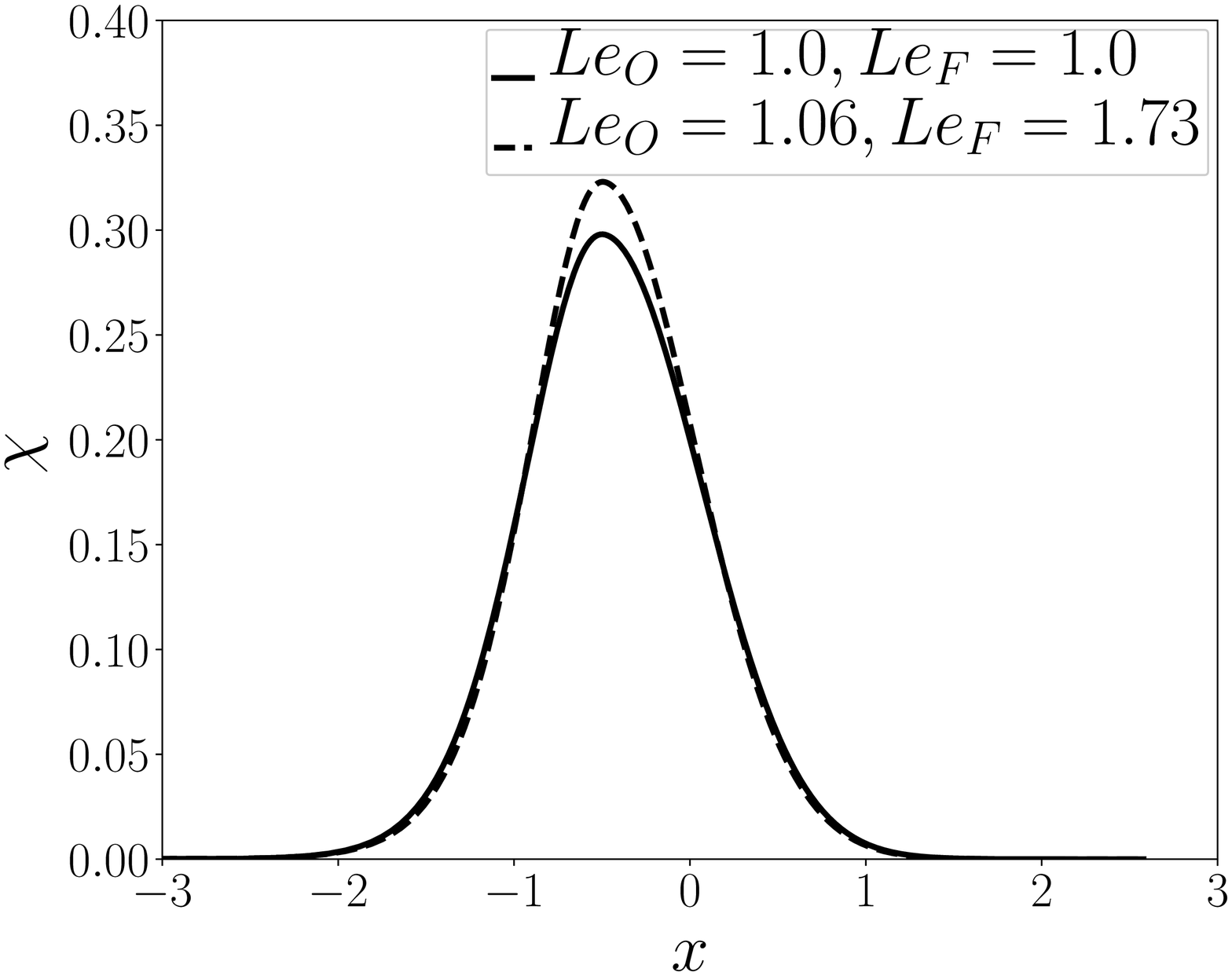}
\includegraphics{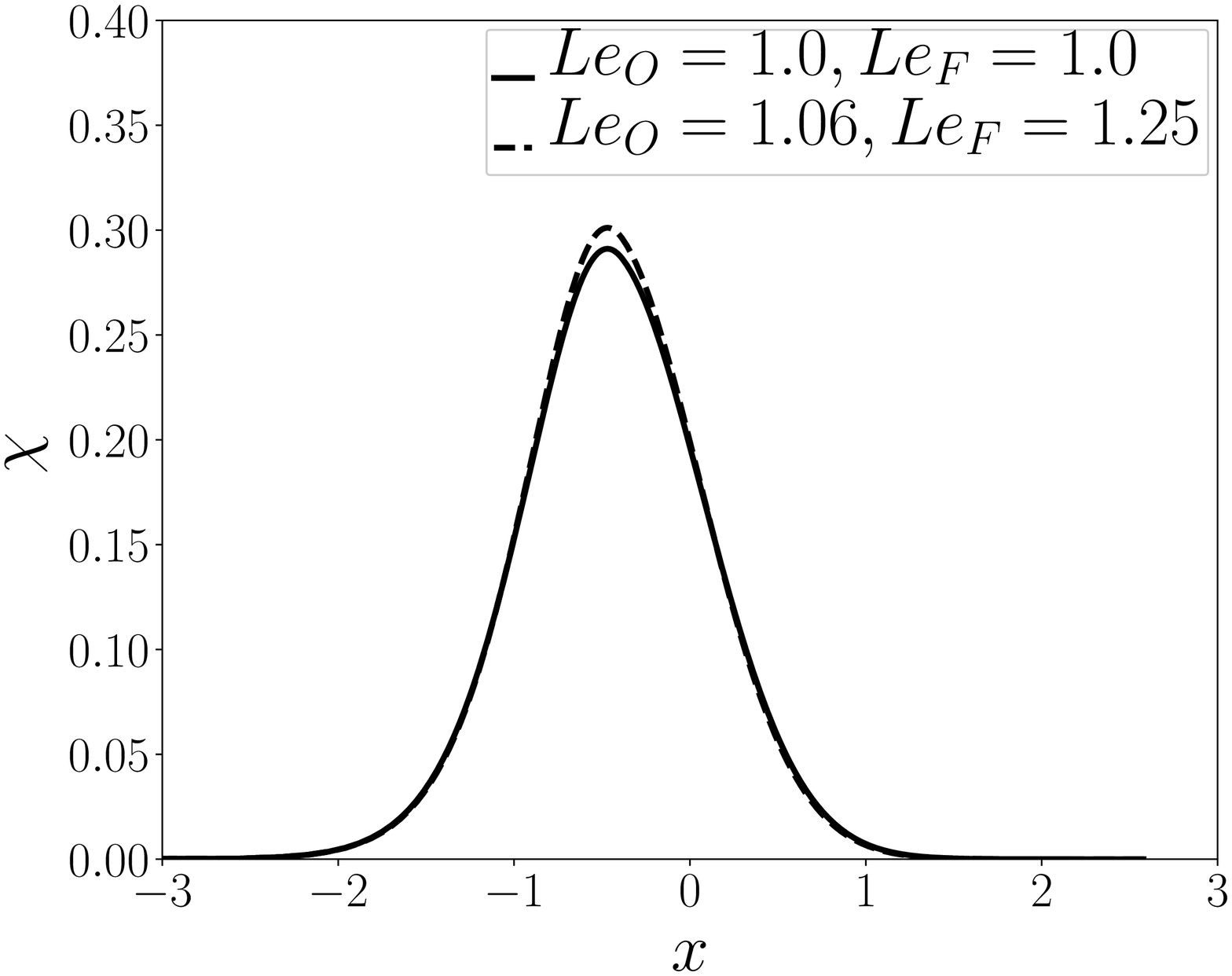}
}
\hspace{0.1\textwidth} (a)\hspace{0.45\textwidth} (b) 
\caption{Comparison of $\chi$ and $\chi_Z$ between unity $Le$ and constant but non-unity $Le$ results for (a) C$_2$H$_6$O (ethanol) and (b) CH$_4$O (methanol).}\label{fig07b}
\end{center}
\end{figure}

A $Le$ number increase results in enhanced heat transfer with respect to mass diffusion.
Consequently, higher values of $Le$ result in improved heat removal from the flame towards the cold reactants sides.
The decrease on the flame temperature leads to a decrease on the vaporisation rate, and thus to smaller values of the gaseous fuel content, $Y_F$, on the fuel side.
The lower gaseous fuel content yields a flame that attains stoichiometric conditions further into the fuel side, as seen in Fig.~\ref{fig07}.
Note that this behaviour is the same for both fuels, since both have Lewis number greater than unity.
%
Additionally,
$\chi_Z$ and $\chi$ profiles in $x$-space are shown in Fig.~\ref{fig07b}. Both scalar dissipation rates, $\chi_Z$ and $\chi$, represent the inverse of the characteristic mass diffusion time \cite{Peters1984}, consequently an increase in the Lewis number directly translates into higher $\chi_Z$ and $\chi$ values (see Fig. ~\ref{fig07b}).


\subsection{The effect of varying $St$}
\label{subsec:St_effects}
The results presented in this section were obtained using the same set of parameters as in the previous subsection but for Stokes number varying from $0 \le St \le 0.1$, its limit value according to Eq.~(\ref{eq13}).

\begin{figure}[h!]
\begin{center}
\resizebox*{\textwidth}{!}{
\includegraphics{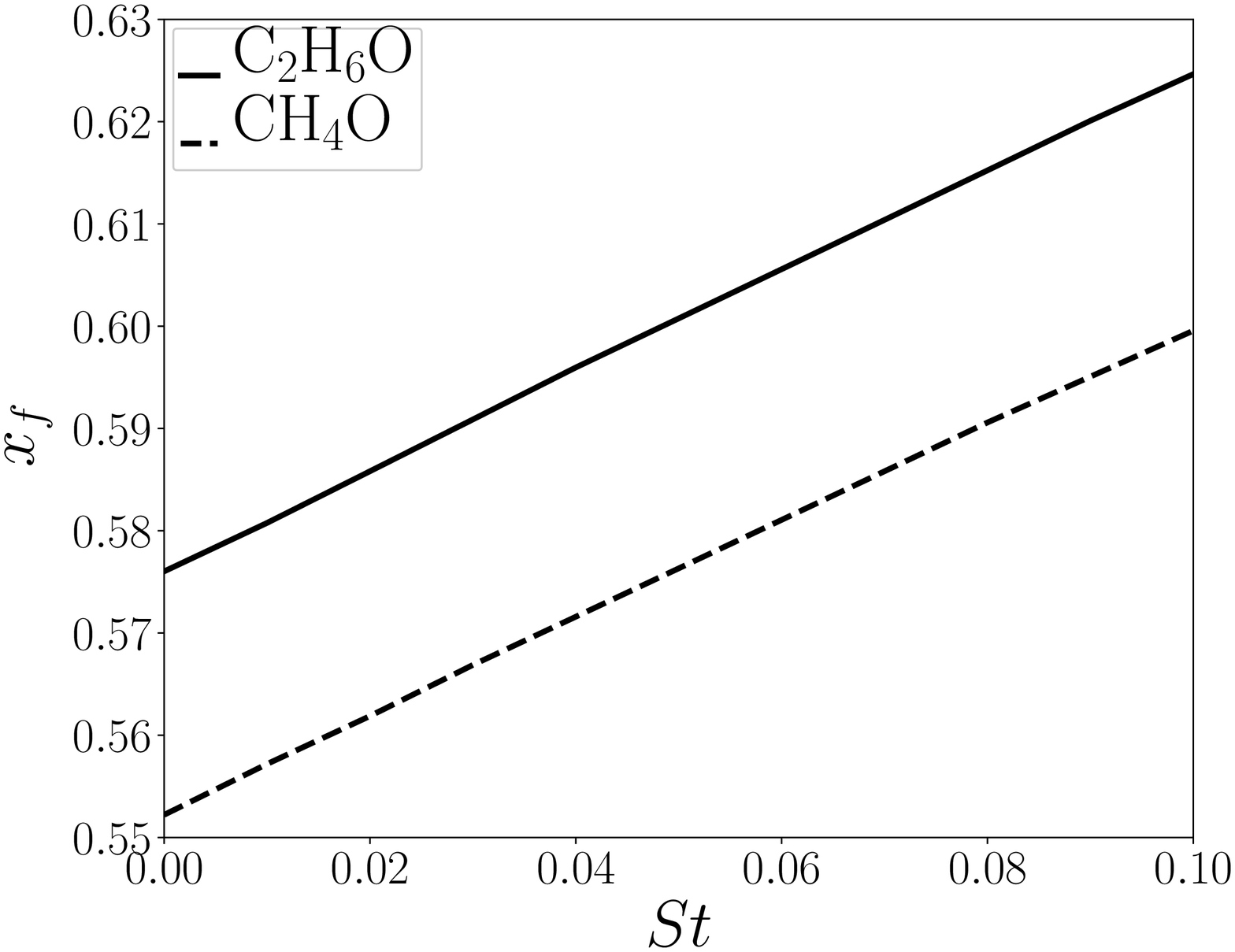}
\includegraphics{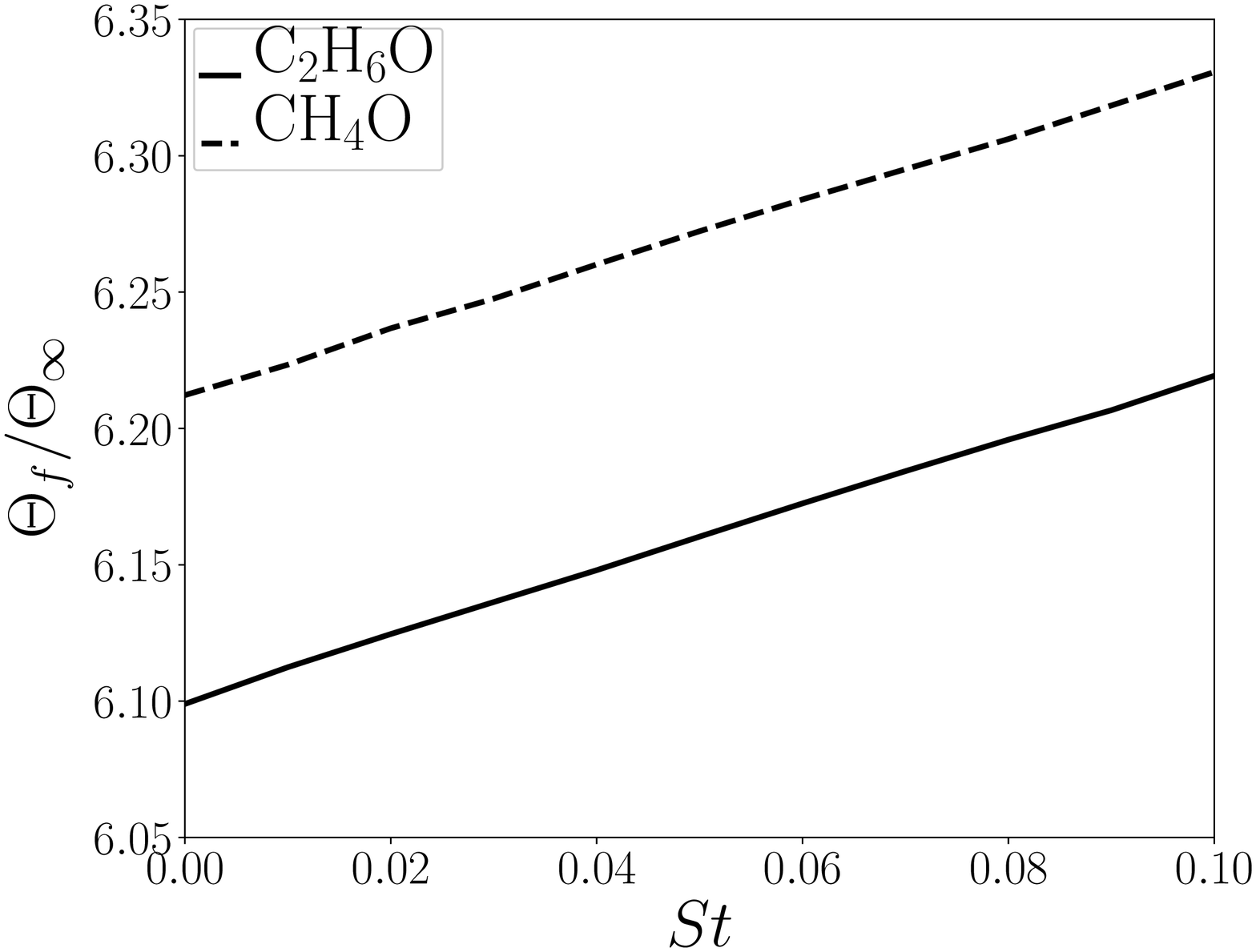}
}
\hspace{0.1\textwidth} (a)\hspace{0.45\textwidth} (b) 
\caption{(a) Flame position $x_f$ and (b) flame temperature $\Theta_f$ as function of the Stokes number for ethanol ($Le = 1.73$) and methanol ($Le = 1.25$) with $Le_O = 1.06$}.\label{fig08}
\end{center}
\end{figure}

\begin{figure}[h!]
\begin{center}
\resizebox*{\textwidth}{!}{
\includegraphics{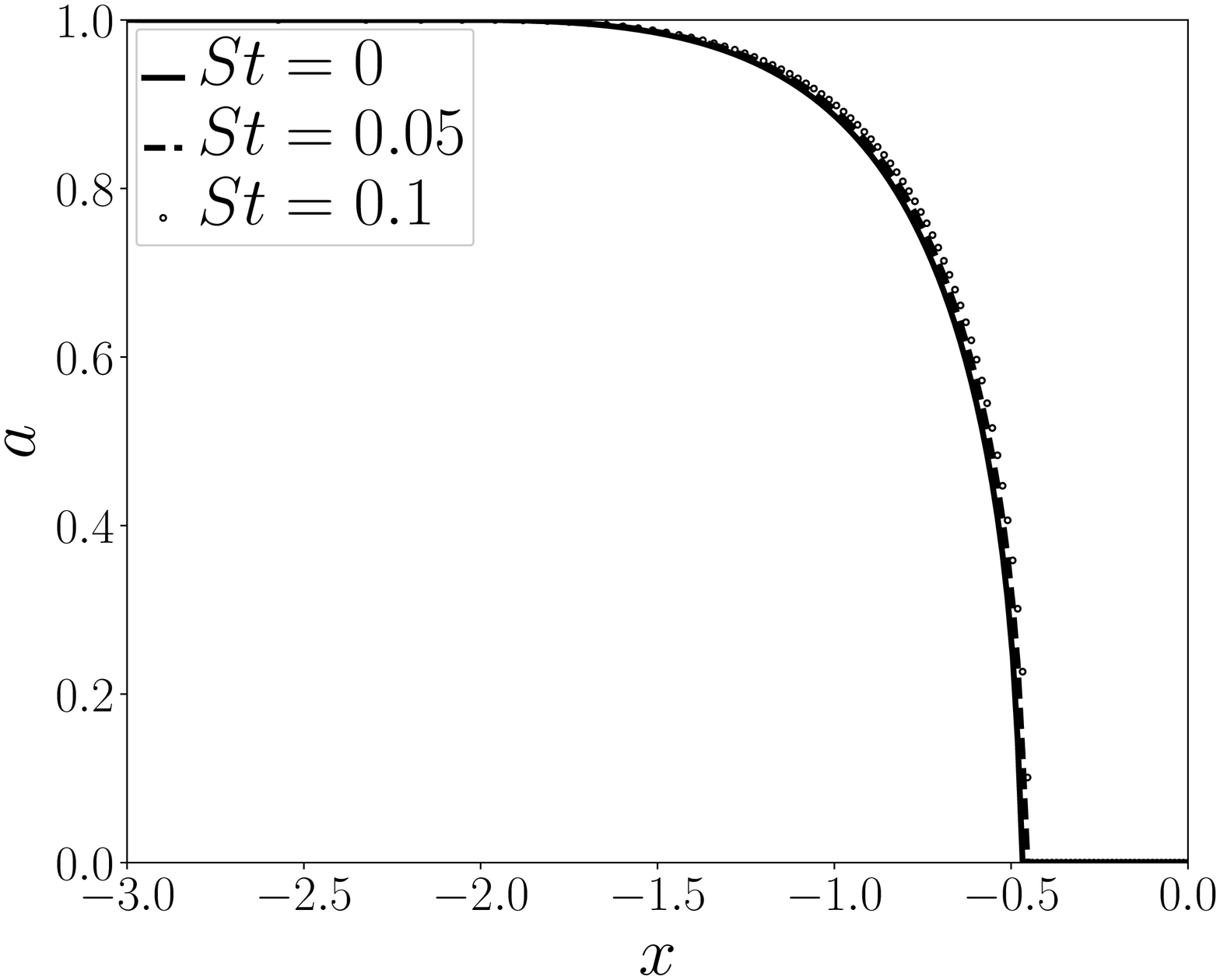}
\includegraphics{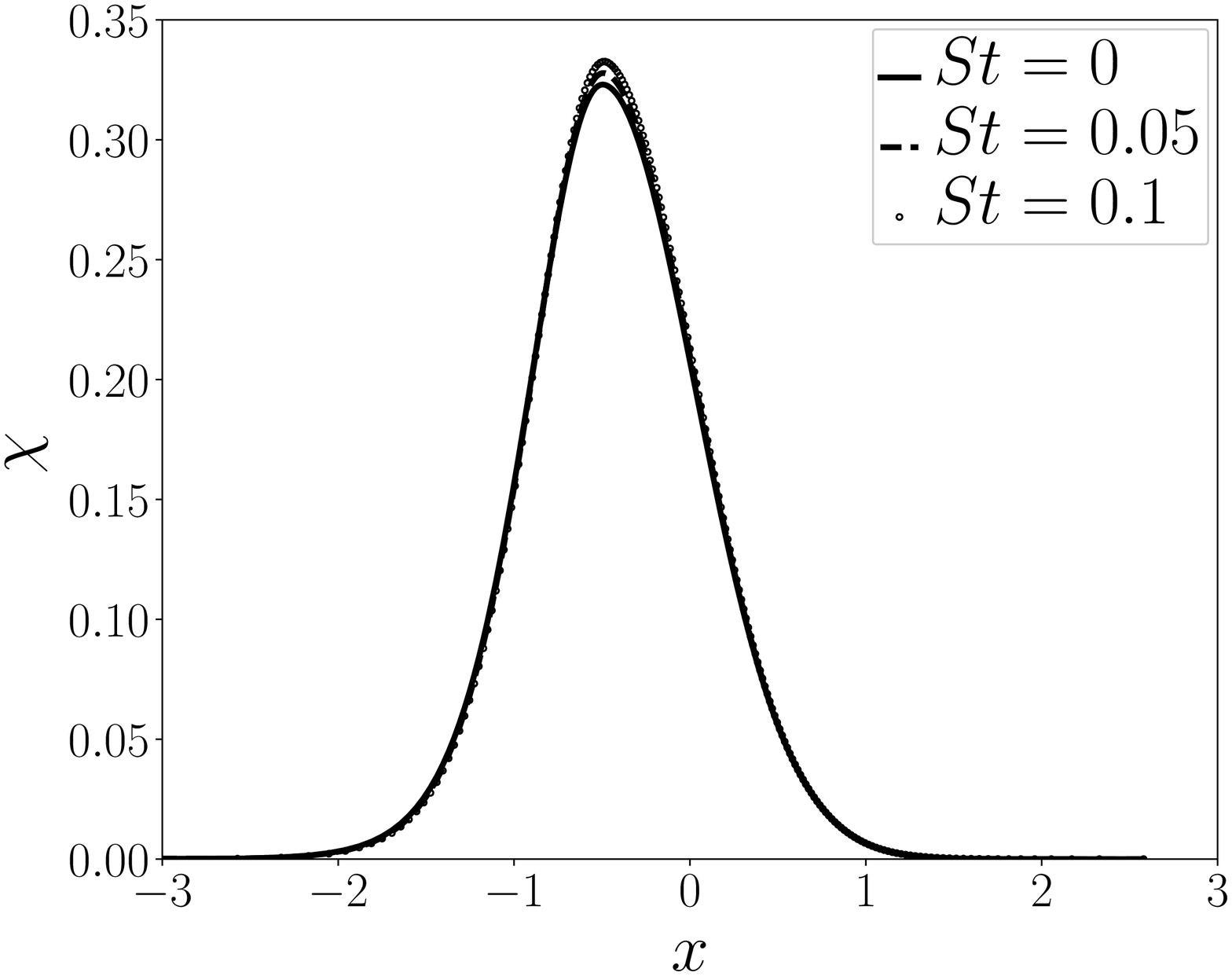}
}
\hspace{0.1\textwidth} (a)\hspace{0.45\textwidth} (b) 
\caption{Profiles of the (a) droplet radius $a$ and (b) dissipation rate $\chi$ for different values of the Stokes number for ethanol ($Le = 1.73$) with $Le_O = 1.06$. The droplets vaporise completely prior to reaching the flame.} 
\label{fig09}
\end{center}
\end{figure}

In Figs.~\ref{fig08} and~\ref{fig09}, the influence of the Stokes number $St$ on the flame position, $x_f$, flame temperature, $\Theta_f$, droplet radius, $a$, and the scalar dissipation rate, $\chi$, are shown.
$a$ and $\chi$ are presented in physical space as it provides a more intuitive picture with $x<0$ being the fuel region, and $x>0$ being the oxidant region.
An increase on the Stokes number $St$ is equivalent to having a larger initial droplet size. 
Since we assume complete combustion, this leads to more fuel reaching the flame, whose end result is that the flame is pushed towards the oxidant side and its temperature increases. 
These effects are clearly seen in Figs \ref{fig08} (a) and (b), respectively.

Note that the droplets vaporize completely prior to reaching the flame (compare Figs.~ \ref{fig08}(a) and \ref{fig09}(a)) because of the assumption of complete combustion;
if it were to be relaxed, unvaporized droplets may reach the flame, leading to heat removal from the reaction zone and a subsequent decrease in flame temperature \cite{gutheil1998}.
Furthermore, if the droplets cross over the stagnation plane towards the incoming oxidant stream, they are subsequently brought back towards the fuel side. 
This flow behavior may lead to oscillations in the flame front, destabilising the flamelet, as discussed in~\cite{greenberg2017}.
Finally, if two-way coupling were to be considered, i.e., that the droplets also affect the gaseous flow field, an increase in $St$ would push the flame even further towards the oxidant side due to droplets inertia \cite{watanabe2007}.

We emphasise that a more realistic evaluation of the influence of the Stokes number in the problem at hand, would unavoidably require accounting for two-way coupling. 
This is nonetheless left for a future investigation.
In the present case, where potential flow was assumed (i.e., the liquid phase does not influence the gaseous phase), changes in $St$ are only brought about through the droplet radius, see Eq.~(\ref{ap04}) in the Appendix.
This effect is small, as seen in Fig.~\ref{fig09}~(a), where the spatial distribution of the droplet radius, $a$, for three values of $St$ is shown.
Since the gas flow field is not affected by the droplets, the changes in the scalar dissipation rate, $\chi$, as a function of $St$ are also small, as seen in Fig.~\ref{fig09}~(b).
While the results in Fig.~\ref{fig09} are only shown for C$_2$H$_6$O, those for CH$_4$O show the same trends.

\section{Conclusions}
The cumulative mixture fraction variable, $Z_C$, was proposed for the description of the spray-flamelet structure in a counterflow configuration accounting for variable $Le$ and $St$ numbers. 
The flamelet formulation was derived and the feasibility of directly integrating the resulting spray-flamelet equations in $Z_C$-space was demonstrated.
Our results show that in contrast to the traditionally used variable for flamelet description, $Z(x_i)$, the cumulative mixture fraction, $Z_C(x_i)$, is a monotonic function, allowing  temperature, $\Theta(Z_C)$, and mass fractions, $Y_O(Z_C)$ and $Y_F(Z_C)$, to be uniquely defined in this space.
Similarly, the scalar dissipation rate, $\chi$, defined in terms of $Z_C$ was also shown to be a smooth function in physical space $x$.
Notably, a closure relation is not required to describe its behaviour because it can be directly obtained from the proposed $Z_C$-space formulation.
The influence of fuel effects, through their $Le$ numbers, and droplet size, through variations in $St$ numbers, were analyzed. 
The spray-flamelet structure was found to be sensitive to increasing $Le$, with the flame stabilizing earlier towards the fuel side. 
In contrast, increasing $St$ favored stabilization towards the oxidant side. These observations are in line with the expected physical behavior.
%
Future work will include testing our formulation as a subgrid model in multi-phase and multi-component turbulent flows.

\section{Acknowledgements}

This work was supported by CNPq Grant No. 474682/2013-7 and FAPEMAT Grant No. 157028/2014.

\section{Appendix - Vaporisation model}\label{ap}

The model for the vaporisation of isolated droplets was developed in a previous work for $St=0$ \cite{Maionchi2013}, and here we extend it to account for $St\neq0$.
For the liquid phase with constant liquid density $\bar{\rho}_l$ we only need to solve for the total mass of liquid \cite{Fachini1999}
\begin{equation}
\frac{\partial}{\partial x_i}(n_l \rho_l V_l u_{li}) = -\alpha_O S_v.  \label{ap01}  
\end{equation}

If we consider that the spatial variation of the droplets volume is much larger than the variation in its velocity, i.e., $n_l\rho_l u_{li}\partial V_l /\partial x_i \gg V_l\partial (n_l\rho_l u_{l_i})/\partial x_i$, Eq. (\ref{ap01}) becomes
\begin{equation}
n_l \rho_l u_{li} \frac{\partial}{\partial x_i}V_l  = -\alpha_O S_v.  \label{ap02}  
\end{equation}
For the 1-D potential flow with spherical droplets
\begin{equation}
    u_l = - x (1 + a^2 St), \ \ \ 
    V_l = \frac{4\pi}{3}a^3,
    \label{ap02b}
\end{equation}
such that the droplets mass conservation is given by
\begin{equation}
    x \frac{da^3}{dx} = -3\lambda_{
    ef}, \ \ \ \lambda_{ef}
    = \frac{\lambda(x)}{1+a^2 St} = \frac{\alpha_0 S_v}{4\pi n_l \rho_l (1+a^2 St)}\label{ap03}
\end{equation}
in which $\lambda$ is the vaporisation rate.
Defining the vaporisation function $\beta = \lambda/a$ and integrating Eq.~(\ref{ap01}), the
droplet radius is given by the nonlinear relation
\begin{equation}
a^2\left(1+\frac{a^2}{2}St\right) = \left[1 + \frac{St}{2} + 2 \int_{-\infty}^{x} \frac{\beta(s)}{s} ds\right]\mathcal{H}(T-T_B), \label{ap04}
\end{equation}
in which $\mathcal{H}$ is Heaviside function.
The vaporisation function $\beta(x)$ depends on the ambient temperature and on the temperature of the droplet (in our case, the liquid fuel boiling temperature,  $T_B$)\cite{Fachini1999}
\begin{equation}
\beta(x) = \int_{T_{B}}^{T_{l\infty}} \frac{dT}{T-T_{B} +L_v} = \ln\left(1+\frac{T_{l\infty} - T_B}{L_v}\right),
\end{equation}\label{ap05}
The ambient temperature for the droplet, $T_{l\infty}$, corresponds to the local temperature in the
spray problem.
These expressions are obtained from the classical model for vaporisation of isolated droplets \citep{Fachini1999,Maionchi2013}. 
From Eq.~(\ref{ap03})  and using $n_l = f_l/(4\pi a^3/3)$, the source term can be written as
\begin{equation}
S_v  =  M \lambda_{ef}, \ \ \ M= \frac{3 Le_O \nu }{Y_{O\infty}} \frac{f_l \rho_l}{a^3} (1+a^2 St)    \label{eq29}  
\end{equation}\label{ap06}
It is worth noting that the spray combustion parameter $M$ combines properties of chemical reaction, flow field and spray \cite{Maionchi2017}.
This non-dimensional parameter permits the analysis of the influence of the initial droplet radius on the spray-flamelet structure by just varying $M$. 
Conveniently, the combustion process can be studied by analysing only this parameter and not the individual influence of each of its constituent parts.

Note that the vaporisation source term, $S_v$, does not depend on $St$ (a consequence of considering a potential flow) and is zero for the following situations: i) $M=0$ if the gas temperature not reached the boiling temperature yet, and ii) $\lambda =0$ if the droplets are totally vaporised. 
%

\section*{References}

\bibliography{Zc}

\end{document}